\documentclass[12pt]{article}
\DeclareUnicodeCharacter{0301}{\'{e}}
\usepackage{amssymb}
\usepackage{amsmath}
\usepackage{amsthm}
\usepackage{color}
\usepackage{comment}
\usepackage{enumitem}		
\usepackage{geometry}		
\usepackage{graphicx}
\usepackage{authblk}
\usepackage{amssymb}
\usepackage{listings}
\usepackage{xcolor}
\usepackage{caption}
\usepackage{color}
\usepackage{algorithm}
\usepackage{algpseudocode}

\usepackage{subcaption}

\makeatletter
	\@addtoreset{equation}{section}
\makeatother

\let\originalleft\left
\let\originalright\right
\renewcommand{\left}{\mathopen{}\mathclose\bgroup\originalleft}
\renewcommand{\right}{\aftergroup\egroup\originalright}

\newlist{romanlist}{enumerate}{3}
\setlist[romanlist]{label=\roman*),ref=(\roman*)}

\begin{document}

\newcommand{\cF}{\mathcal{F}}
\newcommand{\cP}{\mathcal{P}}
\newcommand{\cR}{\mathcal{R}}
\newcommand{\cS}{\mathcal{S}}
\newcommand{\cT}{\mathcal{T}}
\newcommand{\ee}{\varepsilon}
\newcommand{\rD}{{\rm D}}
\newcommand{\re}{{\rm e}}

\newtheorem{theorem}{Theorem}[section]
\newtheorem{corollary}[theorem]{Corollary}
\newtheorem{lemma}[theorem]{Lemma}
\newtheorem{proposition}[theorem]{Proposition}

\theoremstyle{definition}
\newtheorem{definition}{Definition}[section]


\title{
Pathways to hyperchaos in a three-dimensional quadratic map
}

\author[1]{Sishu Shankar Muni}

\affil[1]{School of Digital Sciences,\\ Digital University Kerala\\
Thiruvananthapuram, PIN 695317, Kerala, India}

\maketitle


\begin{abstract}
This paper deals with various routes to hyperchaos with all three positive Lyapunov exponents in a three-dimensional quadratic map. The map under consideration displays strong hyperchaoticity in the sense that in a wider range of parameter space the system showcase three positive Lyapunov exponents. It is shown that the saddle periodic orbits eventually become repellers at this hyperchaotic regime. By computing the distance of the repllers to the attractors as a function of parameters, it is shown that the hyperchaotic attractors absorb the repelling periodic orbits. First we discuss a route from stable fixed point undergoing period-doubling bifurcations to chaos and then hyperchaos, and role of saddle periodic orbits. We then illustrate a route from doubling bifurcation of quasiperiodic closed invariant curves to hyperchaotic attractors. Finally, presence of weak hyperchaotic flow like attractors are discussed.
\end{abstract}

\section{Introduction}
The term ``hyperchaotic" was first coined by R\"{o}ssler when he observed some attractors more complex than chaotic attractors in a nonlinear chemcial model \cite{Rossler79}. Hyperchaotic attractors of a discrete map are chaotic attractors with $n$ positive Lyapunov exponents ($n \geq 2$). For continuous systems, one null Lyapunov exponent along the central manifold combined with one negative Lyapunov exponent for ensuring boundedness of the solution, then for the possibility of existence of hyperchaos, the dimension of the system should be at least $4$. In contrast, for discrete maps, for the possibility of hyperchaos, the dimension of the system should be at least $2$. Much studies  on hyperchaos have been in the case of discrete mapping have been on lower dimensions \cite{UDV}. Many new bifurcations can only occur in higher dimensional maps like for example, doubling bifurcations \cite{Muni23a,ModeLocked23,Muni23b}. In \cite{Shykhmamedov_2023}, bifurcation routes to hyperchaos were considered for a three-dimensional Mir\'{a} map, which is a normal form map of the first return map near homoclinic tangencies of multidimensional systems.
However, in previous studies, three-dimensional phase space along with the saddle periodic points are absent. In this paper, we show three-dimensional phase portraits of hyperchaotic attractors along with the coexisting saddle periodic points of different periods. Moreover, much research and examples of dynamical systems with exactly the same number of positive Lyapunov exponents as the dimension of the system is absent. In this paper we fill that gap and showcase an example of a three-dimensional discrete map  with three positive Lyapunov exponents \cite{Ren_2017} and analyse the dimension and absorbing nature of the unstable invariant periodic orbits inside the hyperchaotic attractor. The unstable periodic orbits are located via the use of multi-dimensional Newton-Raphson method and attractor continuation techniques. The examples presented in the paper gives us a hint about the role played by the doubling bifurcation of both ergodic and resonant torus \cite{Muni23b} in the formation of hyperchaotic attractors.

Hyperchaos was found in coupled higher dimensional systems \cite{KapitiniakRiddling} and the role of transverse direction to the unstable invariant sets were discussed. The relation between the synchronization of chaos and occurence of hyperchaotic attractors were uncovered. Riddling bifurcation refers to the scenario in which one of the unstable periodic orbits absorbed in a chaotic attractor becomes unstable in the direction transverse to the attractor. Such riddling bifurcation leads to the loss of chaos synchronization in coupled identical systems and is a possible mechanism behind the creation of hyperchaotic attractors. A chaos to hyperchaos transition and change in the dimension of the unstable orbits is discussed \cite{Kapit95}. These were later shown in a variety of applications such as in two coupled SQUID timer \cite{SquidHyper21}, where by using the riddling bifurcation, the sysnchronization aspects is studied. In \cite{HyperApp1}, synchronization of complex variable chaotic systems in a discontinuous unidirectional coupling scheme is studied. In \cite{HyperApp2}, the chaos-hyperchaos transition in various models were analysed which are governed by the Sommerfeld effect. In \cite{Shykhmamedov_2023}, a common route to hyperchaos was derived in the sense of doubling of saddle orbits which lead to increment in the dimensionality of their unstable manifolds and such saddle orbits with higher dimensional unstable manifolds were found to be absorbed inside hyperchaotic attractors. On the other hand, importance of the doubling of saddle orbits were seen in \cite{ModeLocked23}, where different types of doubling bifurcations were reported in three-dimensional maps where  doubling of both stable and saddle periodic orbits were observed and saddle-node connections were illustrated via the use of one-dimensional unstable manifolds. Different examples of three-dimensional maps were constructed which illustrates the doubling bifurcation of both mode-locked and ergodic orbits in higher dimensional systems. 

Similar to Shilnikov saddle focus observed in continuous dynamical systems \cite{ShilnikovFocus}, Shilnikov attractors can also exist in higher dimensional maps (the dimension should be atleast three). 
In \cite{Kara21}, simple three-dimensional maps were considered which showcases Shilnikov hyperchaotic attractors in orientation-reversing maps. Prevalence of hyperchaotic attractors and H\'{e}non type hyperchaotic attractors were later explored \cite{Shykhmamedov_2023}. 

Nonlinear systems with hyperchaotic attractors are important in real-world applications such as cryptography, encryption schemes, where more complex behavior than the chaotic behavior is needed. There exists a vast literature which study the applications of hyperchaotic systems. In a four-coupled ring Chua circuit system \cite{KapitiniakChua}, hyperchaos was found. Hyperchaos is evident in fluid mixing and stretching as hyperchaotic attractors occupy a large phase space \cite{FluidHyperchaos}. The dynamical equations governing the propagation of sound waves under the ocean can be reduced to a two-dimensional area-preserving standard mapping \cite{HyperWater}. When the stochasticity parameter related to the environment varies, there are signatures of hyperchaos in the system.  In \cite{Elw99}, an inductorless hyperchaotic generator is discussed. \textcolor{black}{In \cite{Koc92}, experimental demonstration of secure communication via chaotic and hyperchaotic synchronization is discussed. In \cite{Yassen2008}, authors have shown the synchronization of two identical hyperchaotic systems by using active control technique. In \cite{Naderi2016}, authors have discussed exponential synchronization of a chaotic system and also considered its application in secure communication via masking method. In \cite{10374589}, authors have considered hyperchaos based secure communication using Lyapunov theory. In \cite{915393}, authors have experimentally demonstrated observer-based hyperchaos synchronization.}

In \cite{Gao08}, a new image encryption scheme which employs an image shuffling matrix and then uses a system developing hyperchaos. The advantage is of large key space and high security. Hyperchaotic behavior has not been limited to integer order systems, rather the study is explored in fractional order systems and its applications to image encryption \cite{Yang20}. In \cite{Natiq18}, discrete maps exhibiting hyperchaotic behavior is also discussed. A mathematically rigorous cryptanalysis based on hyperchaotic sequences and its application to image encryption is explored in \cite{Ozkaynak12}. Color image encryption algorithms applied to security analysis through the lenss of hyperchaotic systems is studied in \cite{Li19}. To enhance the performance of image encryption, a three-dimensional hyperchaotic map is proposed and further applied to encryption in \cite{Hu23}. Similar performance analysis of discrete memristor hyperchaotic systems and its design were studied in \cite{Lai23a} and its image encryption application in 
\cite{Lai23b}. Realworld hardware realizations of memristive hyperchaotic maps were explored in \cite{Wang23}. This motivated to generate various maps with hyperchaotic behavior \cite{Zhang23,Liu23}. \textcolor{black}{Discrete maps have been realized in electronic circuits \cite{GarcaMartnez2013} in one-dimensional maps. An electronic circuit of 
Poincar\'e map of chemical dynamical system was studied in \cite{Huang2005} where chaoticity is studied using concept of entropies.}\\
The main contributions of this paper are as follows
\begin{itemize}
    \item Identify various routes that leads to all three positive Lyapunov exponents.
    \item Investigate the type of eigenvalues (dimensionality of the saddles)  of mediating unstable invariant objects absorbed in the hyperchaotic attractor.
    \item Exploring the flow like weak hyperchaotic attractor with positive near zero second Lyapunov exponent.
\end{itemize}
The paper is organized as follows. In \S \ref{sec:ShilnikovHyper}, a three-dimensional generalised H\'{e}non map is discussed with hyperchaotic Shilnikov attractors. In \S \ref{sec:twoparam}, we illustrate a two-dimensional color coded Lyapunov chart along with MATCONTM continuation diagram. In \S \ref{sec:stablefp_hyperchaos}, a route to the presence of three positive Lyapunov exponents is shown. In \S \ref{sec:DoublingHyperchaos}, a route from doubling bifurcation of quasiperiodic closed invariant curves to hyperchaos is discussed. In \S \ref{sec:weak_hyperchaos}, presence of weak flow like hyperchaotic attractors are discussed. 

Before discussing the case of three positive Lyapunov exponents, we briefly discuss the case of two positive Lyapunov exponents and Shilnikov hyperchaotic attractor in a 3D generalised H\'{e}non map.

\section{From resonant torus doubling to Shilnikov hyperchaotic attractor}
\label{sec:ShilnikovHyper}
In the previous studies \cite{Kara21}, Shilnikov hyperchaotic attractors were shown in three-dimensional orientation-reversing maps. In this section, we introduce a generalised H\'{e}non map in three dimensions with constant Jacobian which allows us to make the map orientation-reversing or orientation-preserving depending upon the sign of a single parameter value. The three-dimensional Generalised H\'{e}non map considered in \cite{ModeLocked23} is given by 
\begin{equation}
    \begin{aligned}
    x_{n+1} &= a - y_{n}^2 - bz_{n},\\
    y_{n+1} &= x_{n},\\
    z_{n+1} &= y_{n},
    \end{aligned}
    \label{eq:HyperGenHenmap}
\end{equation}
where $x,y,z$ represent the state variables, and $a,b$ are parameters. The determinant of the Jacobian of the map \eqref{eq:HyperGenHenmap} is $-b$ and thus for $b>0$, the map is orientation-reversing everywhere, and for $b<0$, the map is orientation-preserving everywhere. Note that map \eqref{eq:HyperGenHenmap} is invertible. If say $(x,y,z)$ maps to $(x',y',z')$ under $f$, then $f^{-1}$ is given by 
\begin{equation}
    \begin{aligned}
    x  &= y',\\
    y &= z',\\
    z  &= \frac{a-z'^2 - x'}{b}.
    \end{aligned}
    \label{eq:InverseHyperGenHenmap}
\end{equation}
\begin{figure}[!htbp]
\begin{center}
\includegraphics[width=0.4\textwidth]{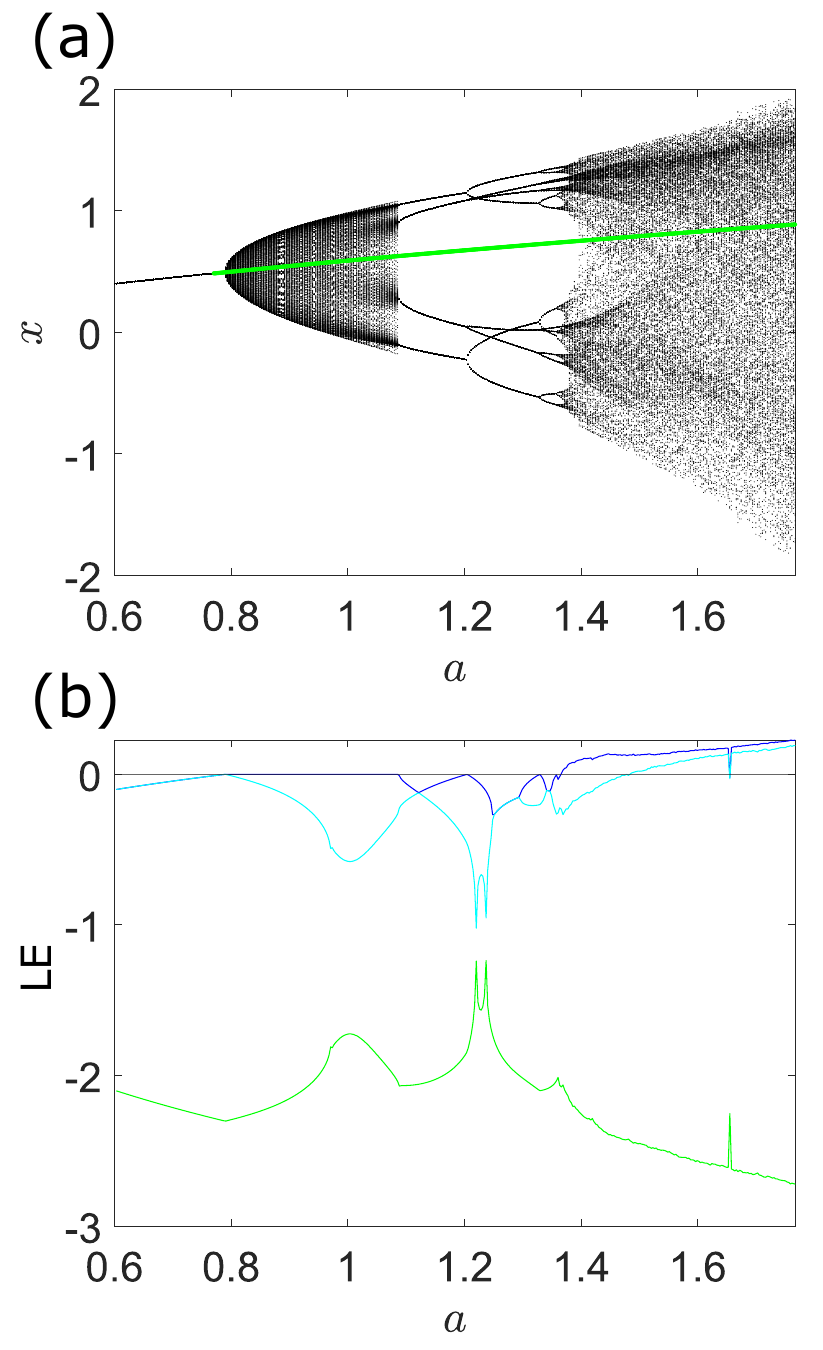}

\end{center}
\caption{In (a) a one-parameter bifurcation diagram of $x$ vs $a$ is shown along with the continuation of saddle fixed point (in green). In (b), one-parameter variation of the three Lyapunov exponents with parameter $a$.}
\label{fig:HyperGenHen_OneParam_Saddles}
\end{figure}
A one-parameter bifurcation diagram of $x$ vs $a$ is shown in Fig. \ref{fig:HyperGenHen_OneParam_Saddles} (a) along with variation of Lyapunov exponents in Fig. \ref{fig:HyperGenHen_OneParam_Saddles} (b). \textcolor{black}{We have used QR method \cite{QRmethod} to compute the Lyapunov exponent spectrum of the 3D map. At each parameter value, an orbit is selected within an attractor and the exponents are assessed over $10^6$ iterations. For each point in the parameter interval, the map is iterated starting from a random initial condition and Lyapunov spectrum is computed}.
We also continuate the saddle fixed point shown in green dots. It was shown in \cite{Muni23b} that for $a > 1.1$, the map exhibits resonant torus doubling bifurcation. It can be observed in Fig. \ref{fig:HyperGenHen_OneParam_Saddles} (a) that after subsequent period doubling bifurcation, the system enters to regime of chaos and further makes a transition to hyperchaos with 
two positive Lyapunov exponents, see Fig. \ref{fig:HyperGenHen_OneParam_Saddles} (b). 

\begin{figure*}[!htbp]
\begin{center}
\includegraphics[width=0.8\textwidth]{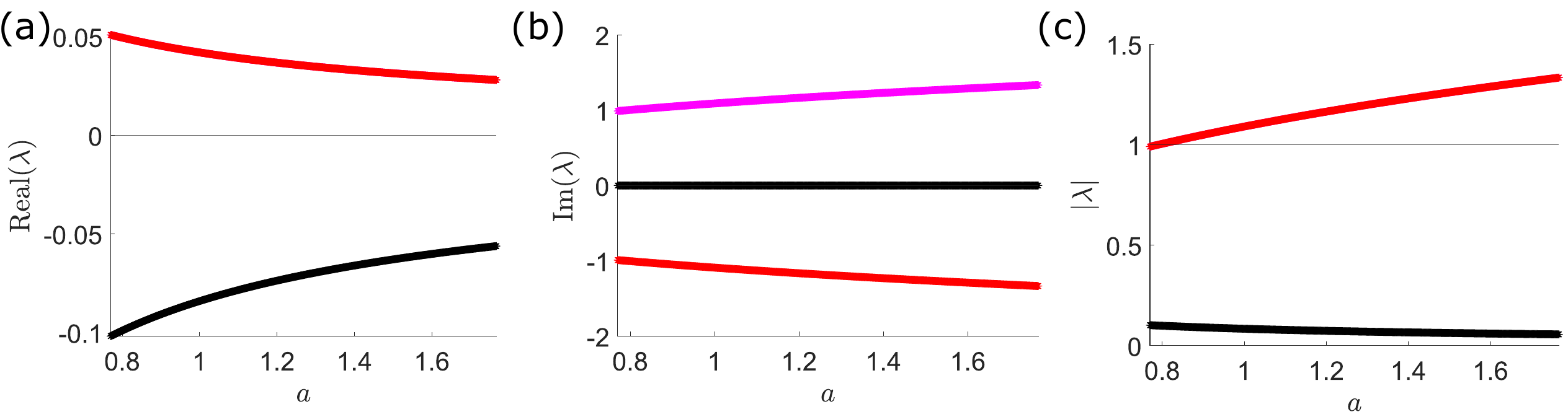}

\end{center}
\caption{Variation of the eigenvalues of saddle fixed point with parameter $a$. In (a), (b), (c) the real, imaginary, and absolute part of the eigenvalue of the saddle fixed point is shown. Observe the saddle-focus nature of the saddle fixed point with increase in parameter $a$.}
\label{fig:HyperGenHenEigen}
\end{figure*}

\begin{figure*}[!htbp]
\begin{center}
\includegraphics[width=0.8\textwidth]{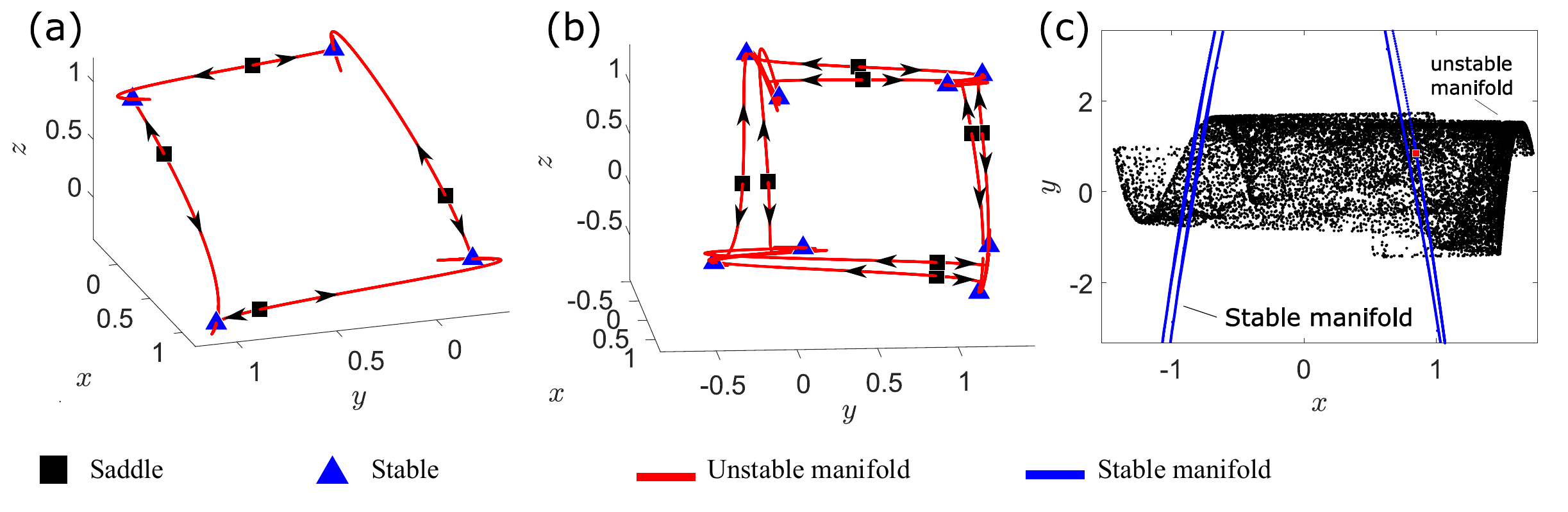}
\end{center}
\caption{In (a), a resonant-mode locked period-four orbit is shown for $a = 1.2$ with the saddle-node connection formed by the 1D unstable manifolds (in red). In (b), a length doubled resonant mode-locked period-eight orbit is shown for $a = 1.3$. In (c), a Shilnikov hyperchaotic is shown with transversal intersections of 1D stable manifolds (in blue) and 2D unstable manifolds for $a = 1.65$. Parameter $b = 0.1$ is fixed.}
\label{fig:ResonantHyperchaos}
\end{figure*}

In Fig. \ref{fig:HyperGenHenEigen}, we continuate the eigenvalues of the saddle fixed point with parameter $a$. In (a), (b), and (c), the real part, imaginary part, and absolute values of the eigenvalues of the saddle fixed point are shown. Observe that after the fixed point undergoes a supercritical Neimark-Sacker bifurcation for $a = 0.8$, the fixed point develops two complex conjugate eigenvalues with modulus greater than one and third eigenvalue is negative with modulus less than one. 

A saddle fixed point is the simplest unstable invariant set absorbed inside the hyperchaotic attractor. Thus we discuss the absorption and type of the saddle fixed point in the hyperchaotic attractor. The saddle fixed point is of type saddle-focus with two-dimensional unstable manifold and a one-dimensional stable manifold with negative eigenvalue. For $a > 1.5$, the regime of hyperchaos appears and the saddle fixed point of type saddle-focus is absorbed inside the hyperchaotic attractor. Thus the hyperchaotic attractor is a Shilnikov type hyperchaotic attractor. 

In Fig. \ref{fig:ResonantHyperchaos}, we show phase portraits associated with the formation of hyperchaotic attractors. In Fig. \ref{fig:ResonantHyperchaos} (a), we show a mode-locked period-four orbit where the saddle-node connection is formed by the one-dimensional unstable manifold (in red) of the saddle points. In Fig. \ref{fig:ResonantHyperchaos} (b), we show a doubled mode-locked period-eight orbit. In Fig. \ref{fig:ResonantHyperchaos}(c), we show the Shilnikov hyperchaotic attractor with the saddle fixed point (in red square) and its corresponding one-dimensional stable manifold. Observe that transversal intersection by the one-dimensional stable manifold and two-dimensional unstable manifold (in black) of the saddle-fixed point absorbed in the Shilnikov type hyperchaotic attractor. 

The next section focuses on a 3D discrete map \cite{Muni23b} which exhibits all three positive Lyapunov exponents.  We first detect the region of hyperchaos via the two-parameter Lyapunov charts.
\section{Two-parameter diagram to detect hyperchaotic attractors}
\label{sec:twoparam}
\begin{figure*}[!htbp]
\begin{center}
\includegraphics[width=0.8\textwidth]{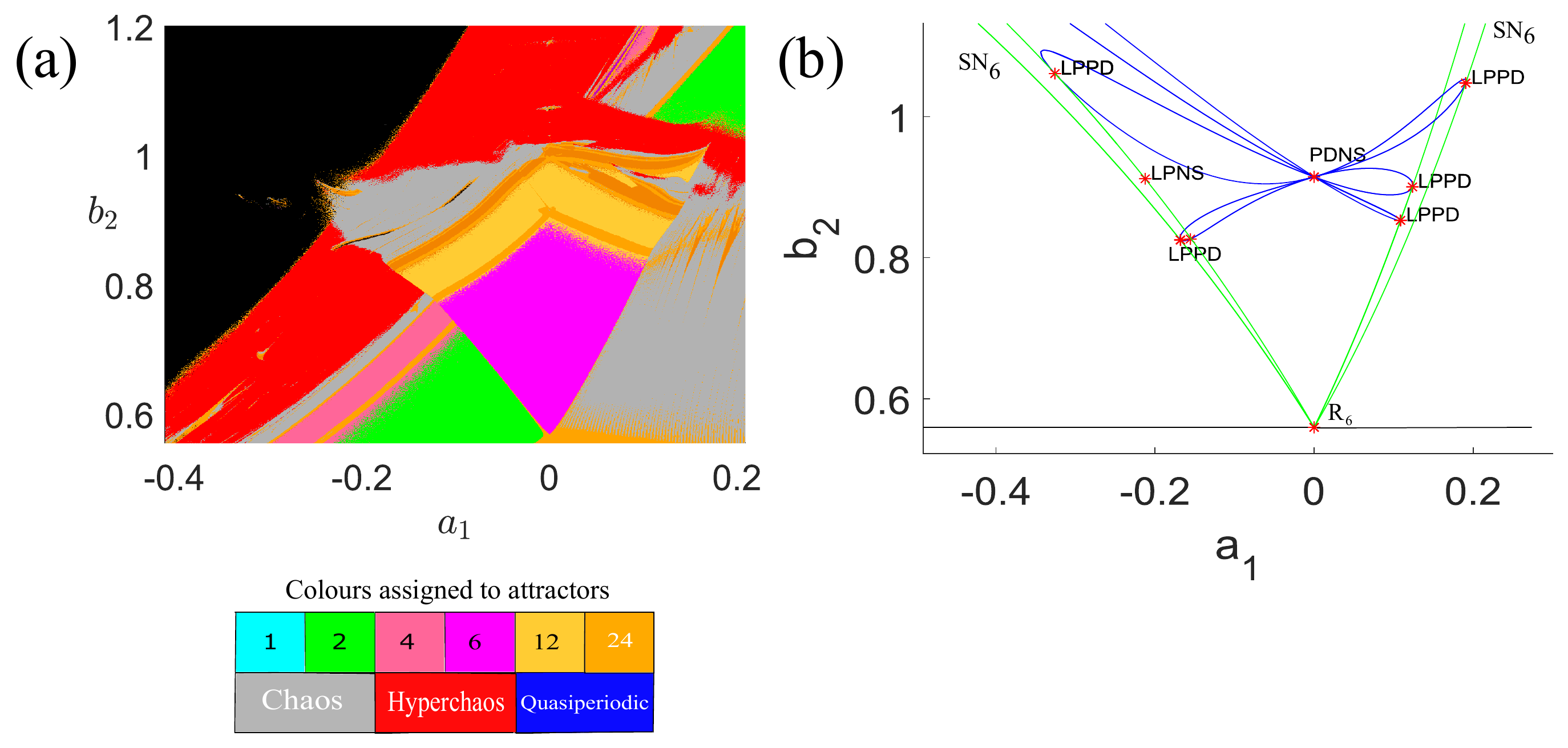}
\end{center}
\caption{In (a), a two-parameter $a_1-b_2$ Lyapunov diagram is shown. Different attractors are color coded according to the legend. The arrow shows a slice for the transition from resonance region to hyperchaos. In (b), corresponding detailed two-parameter bifurcation diagram is shown via the use of continuation techniques using \sc{MATCONTM}. }
\label{fig:MatcontmBasins}
\end{figure*}
The three-dimensional quadratic map \cite{Ren_2017} motivated via a chaos control method is represented as 
\begin{equation}
    \begin{aligned}
    x_{n+1} &= a_{1}x_{n} + a_{2}y_{n} + a_{3}y_{n}^2,\\
    y_{n+1} &= b_{1} - b_{2}z_{n},\\
    z_{n+1} &= cx_{n},
    \end{aligned}
    \label{eq:STMmap}
\end{equation}
where $x,y,z$ represents the state variables, and 

$a_{1}, a_{2}, a_{3}, b_{1}, b_{2}, c$ are parameters. In contrast to \eqref{eq:HyperGenHenmap}, the map is non-invertible and does not have a simple parameter independent fixed point. However, many rich dynamics such as ergodic and resonant torus doublings were reported \cite{Muni23b}.

In Fig. \ref{fig:MatcontmBasins}, two-parameter Lyapunov diagram near the $1:6$ resonance occuring at $(a_1,b_2) = (-0.0008,0.577)$ and its corresponding two-parameter diagram is obtained via continuation techniques through MATCONTM \cite{KuMe19}. Different kinds of attractors are color coded according to the legend. For fixed values of $a_1$, over a large parameter interval, as $b_2$ increases, we observe the transition to regime of hyperchaos (denoted in red). We select a slice in the $a_1-b_2$ parameter plane for a fixed value of $a_1$ and study the emergence of hyperchaotic attractor with three positive Lyapunov exponents. We denote the tip of the Arnold tongue corresponding to the $1:6$ resonance as $R_6$. The left and right curves denote the saddle-node bifurcation  (denoted by $SN_{6}$). A similar tongue slightly displaced corresponds to the coexisting stable period-two orbit, see \cite{Muni23b}. Observe that these saddle-node bifurcation curves meet at $R_6$ which denote a codimension-two point (occurence of saddle-node and Neimark-Sacker bifurcation). The blue curves denote the period-doubling bifurcations of the saddle and stable period-six orbit. The curves meet the $SN
_6$ curves at a fold-flip bifurcation (LPPD). Thus at such $(a_1,b_2)$ point, the periodic orbit would have eigenvalue tuple of $(+1,-1)$. Identification of these bifurcation curves are essential in understanding the structure of the two-dimensional Lyapunov diagram.  Observe that the regime of hyperchaos lies above the regime of stable fixed point regime. The next section deals with the route from stable fixed point to a regime of hyperchaos with all three positive Lyapunov exponents.

\section{From stable fixed point to hyperchaos}
\label{sec:stablefp_hyperchaos}
\begin{figure}[!htbp]
\begin{center}
\includegraphics[width=0.9\textwidth]{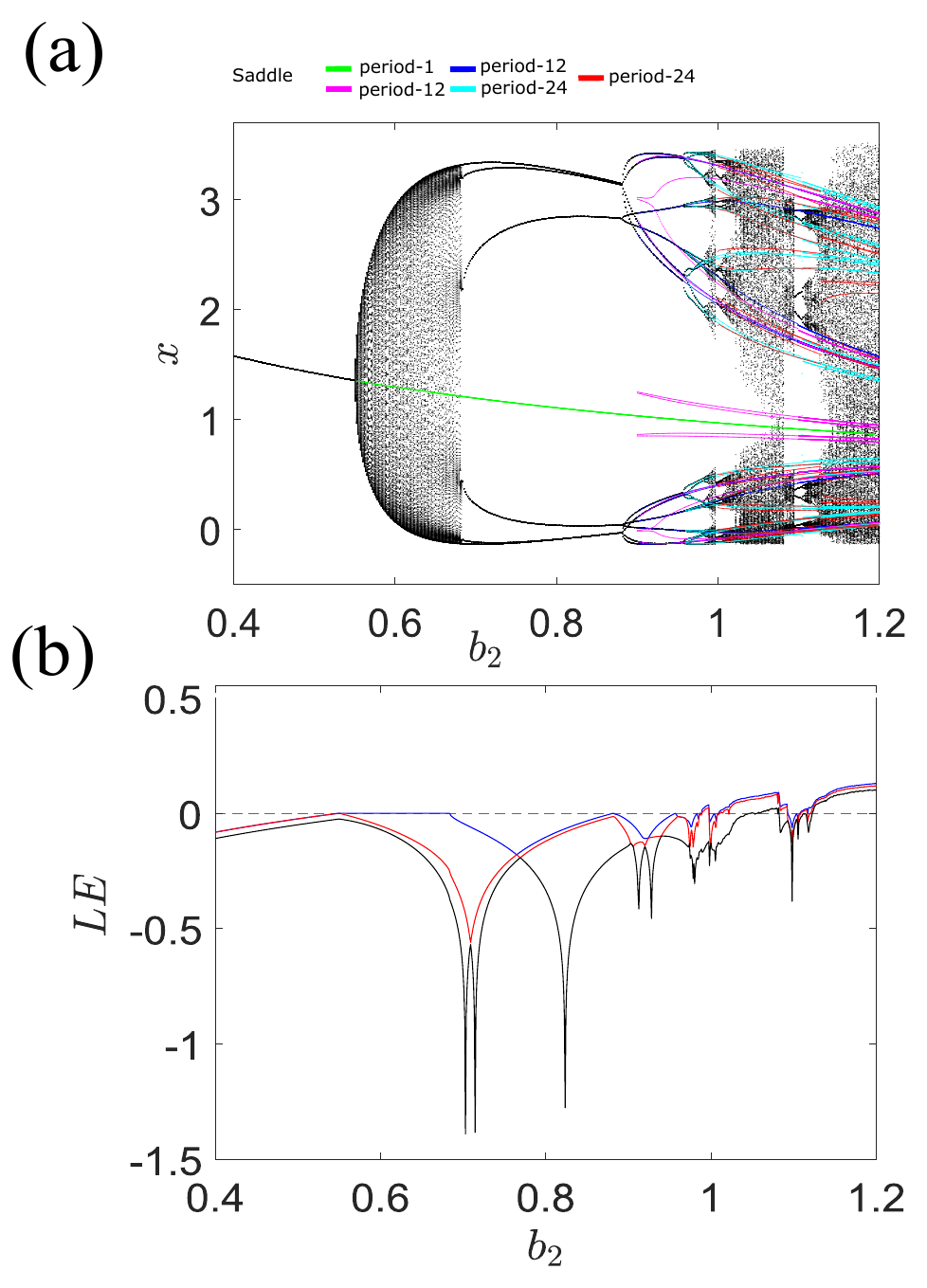}

\end{center}
\caption{In (a) a one-parameter bifurcation diagram of $x$ vs $b_2$ is shown along with the continuation of different saddle periodic orbits distinguishable via their colour as per the legend. In (b), one-parameter variation of the three Lyapunov exponents with parameter $b_2$.}
\label{fig:STM_OneParam_Saddles}
\end{figure}

In Fig. \ref{fig:STM_OneParam_Saddles}, at $b_{2} = 0.4$, there exists a stable fixed point of \eqref{eq:STMmap}. The stable fixed point undergoes a supercritical Neimark-Sacker bifurcation as parameter $b_{2}$ increases.  The quasiperiodic closed invariant curve gets destroyed by a saddle-node bifurcation on invariant circle and formation of a mode-locked period-six orbit takes place. As parameter $b_2$ increases, the stable period-6 orbit undergoes a period-doubling bifurcation. This regime is studied in \cite{Muni23b}, which showcases the doubling bifurcation of resonant mode-locked periodic six orbit. After successive period-doubling bifurcations, system \eqref{eq:STMmap} enters the regime of hyperchaos, as evident from the Lyapunov exponent spectrum shown in Fig. \ref{fig:STM_OneParam_Saddles} (b). For $b_2 \in (0.96,1.2)$, the hyperchaotic regime has 
two positive Lyapunov exponents. This hyperchaotic regime persists with increase in $b_2$, illustrating a robust hyperchaotic regime. The hyperchaos is destroyed by a crisis.

Further increase in $b_{2}$ leads to a regime of hyperchaos in which all three Lyapunov exponents are positive, see Fig. \ref{fig:STM_OneParam_Saddles} (a), (b) for $b_{2}>1.15$. This hyperchaotic regme persists as $b_2$ is increased and thus displays a robust hyperchaotic regime with all three Lyapunov exponents positive. 
The hyperchaotic attractors should have unstable invariant sets with dimension of unstable manifold $\geq 2$ for three-dimensional maps. To understand this behavior in the context of two regimes: (a) hyperchaotic attractor with two positive Lyapunov exponents, and (b) hyperchaotic attractor with three Lyapunov exponents, we continuate various saddle periodic orbits which were previously stable periodic orbits and undergone bifurcations and also saddle orbits coexisting with previous stable orbits forming resonant torus. 

In Fig. \ref{fig:STM_OneParam_Saddles}, we show various saddle periodic orbits in different colour. A saddle fixed point is marked in green, doubled saddle period-12 orbit is marked in magenta, another saddle period-12 orbit in blue, and two different saddle period-24 orbits is marked in red and cyan. As we can see they are continuated till $b_2 = 1.2$, which covers the regime of hyperchaotic attractors. The absorption of the saddle periodic orbits inside the hyperchaotic attractors is not intuitive just via pictorial representation of one-parameter continuation of the saddle periodic points and hyperchaotic attractors. In a similar spirit of \cite{Shykhmamedov_2023}, motivated, we compute the distance between saddle points and hyperchaotic attractor. Minimum Euclidean distance between the saddle periodic point and hyperchaotic attractor is considered for a fixed parameter value. 

\begin{figure*}[!htbp]
\hspace{2cm}\begin{center}
\includegraphics[width=1\textwidth]{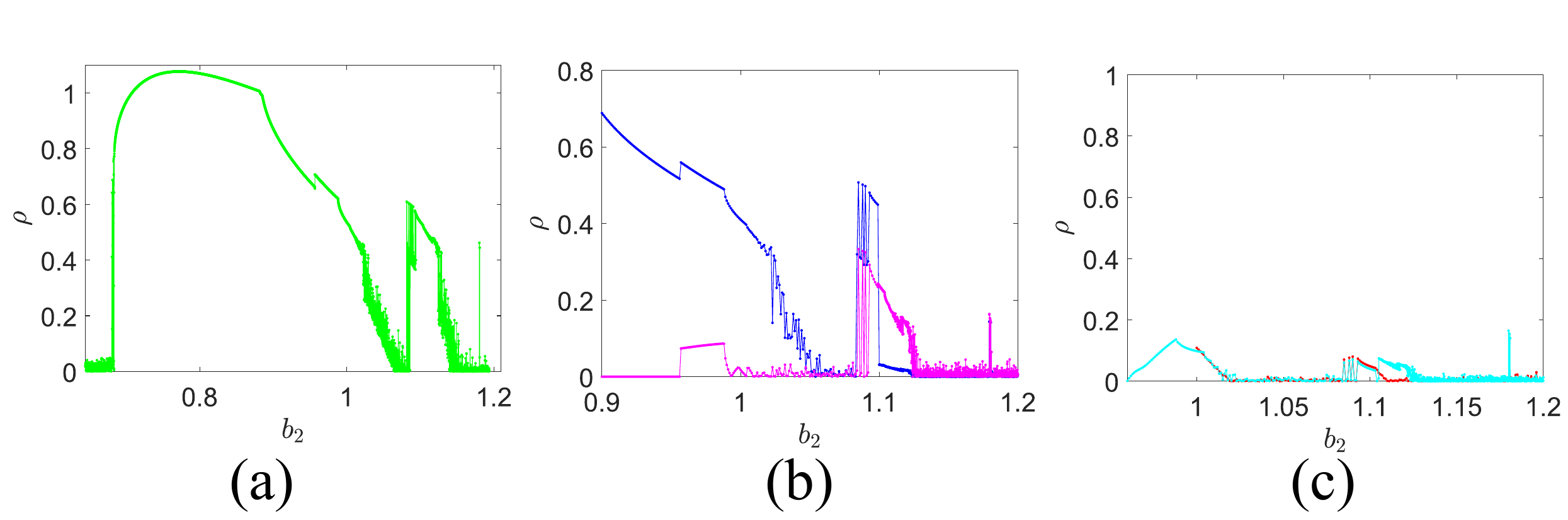}

\end{center}
\caption{Variation of the distance between the saddle fixed points, periodic points and the coexisting attractors. In (a), distance between the unstable fixed point and coexisitng attractors are shown in green. In (b), distance between unstable period-12 orbits (in blue and magenta) and coexisting attractors marked in blue and magneta respectively. In (c), distance between unstable period-24 orbits and the coexisting attractors are shown in red and cyan respectively. Parameter values at which the distance approaches zero suggest that the unstable invariant set is absorbed by the coexisting attractor at that parameter value. }
\label{fig:Distance_Attractors}
\end{figure*}

In Fig. \ref{fig:Distance_Attractors} (a), we show the distance $\rho$ between the saddle fixed point and the attractors in black as in the one-parameter bifurcation diagram in Fig. \ref{fig:STM_OneParam_Saddles}(a). Note that the
distance between the saddle fixed point and the attractors are non-zero initially, but as parameter $b_2$ increases, we observe that the distance between the saddle fixed point and the hyperchaotic attractors are zero (the saddle fixed point is absorbed in the hyperchaotic attractor), see for $b_2>1$. 

In Fig. \ref{fig:Distance_Attractors} (b), distance $\rho$ between two different coexisting saddle period-12 orbits and 
corresponding attractors are shown with the variation of parameter $b_2$. 
For $b_2 < 1.1$, the distance shown in magenta is zero and is mainly due to the fact that the saddle period-12 orbit is absorbed in the quasiperiodic orbit. We note that for $b_2<1.1$, we observe many values of $b_2$ for which the distance is zero, but our main objective is to look for the absorption of the saddle periodic orbits with the hyperchaotic attractors which exists for $b_2 > 1.15$. Note that both the distance marked by magenta and blue are zero, which shows that both the saddle period-12 orbits are absorbed inside the hyperchaotic attractors. 
In Fig. \ref{fig:Distance_Attractors}(c), the distance $ \rho$ between two different coexisting saddle period-24 orbits and corresponding attractors are shown with the variation of parameter $b_2$ in cyan and red dots are shown. Since the hyperchaotic attractors are formed after a period-doubling cascade of the latter periodic orbits, so  we can see that the distance $\rho$ is zero for a greater range of $b_2$ values and more prominent for $b_2 > 1.15$ (the regime of hyperchaotic attractors). This suggests that the saddle period-24 orbits are absorbed in the hyperchaotic attractors. 

This motivates us to illustrate the phase portraits at the parameter values of $b_2$ where the hyperchaotic attractors coexists along with the coexisting saddle periodic orbits. 

Eigenvalues of unstable invariant sets absorbed in the hyperchaotic attractor are 
an important aspect in classifying the type of hyperchaotic attractor and in turn can provide insights with the rich dynamics. For example, the presence of 
saddle-focus orbit with two dimensional unstable manifold absorbed in a hyperchaotic attractor confirms it to be a Shilnikov attractor, with transversal intersections between the two-dimensional unstable manifold and one-dimensional stable manifold.  Motivated, we present the variation of eigenvalues of the saddle fixed point and different saddle periodic orbits shown in Fig. \ref{fig:STM_OneParam_Saddles} (a) as a function of parameter $b_2$. In particular, we present the real part, imaginary part, and the magnitude 
of the eigenavalue for each of the saddle periodic orbits. 
\begin{figure*}[!htbp]
\hspace{2cm}\begin{center}
\includegraphics[width=0.9\textwidth]{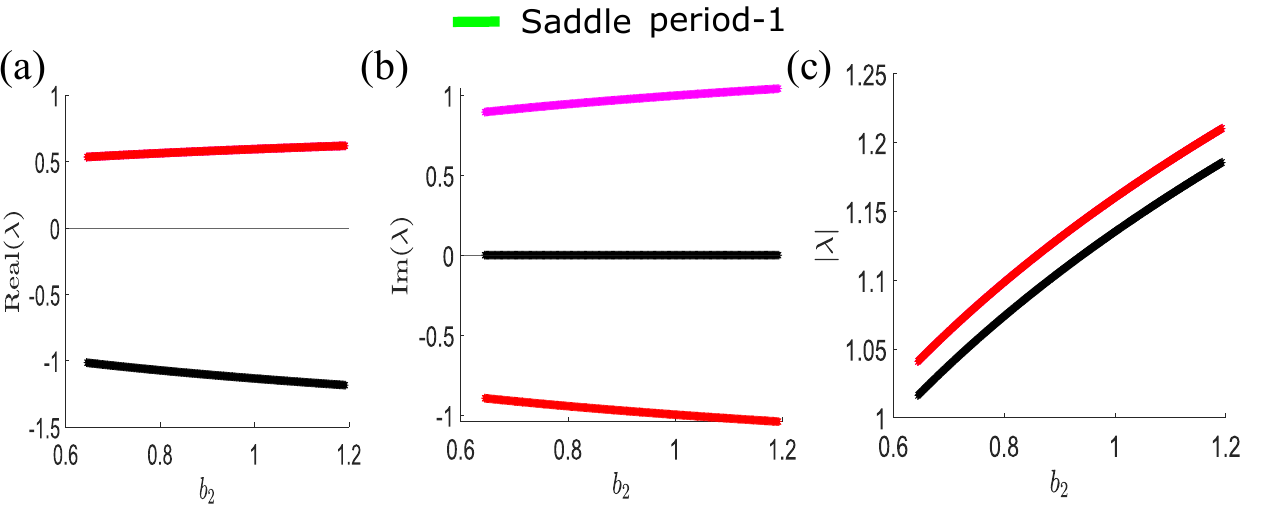}

\end{center}
\caption{One-parameter variation of the eigenvalues of the unstable fixed point with parameter $b_2$. In (a), real part of the three eigenvalues of the unstable fixed point is shown. In (b), imaginary part of the three eigenvalues is shown. In (c), magnitude of the three eigenvalues is considered. }
\label{fig:Eigen_Saddlep1}
\end{figure*}

In Fig. \ref{fig:Eigen_Saddlep1}, we note that for the range of parameter $b_2$ considered, 
the saddle fixed point of \eqref{eq:STMmap}, develops complex conjugate eigenvalues with modulus greater than one as the imaginary part is non-zero and a real eigenvalue with modulus greater than one. Especially for $b_2 > 1.1$, three eigenvalue with modulus greater than unity is observed. The unstable invariant set is a repelling focus.
\begin{figure*}[!htbp]
\hspace{2cm}\begin{center}
\includegraphics[width=0.9\textwidth]{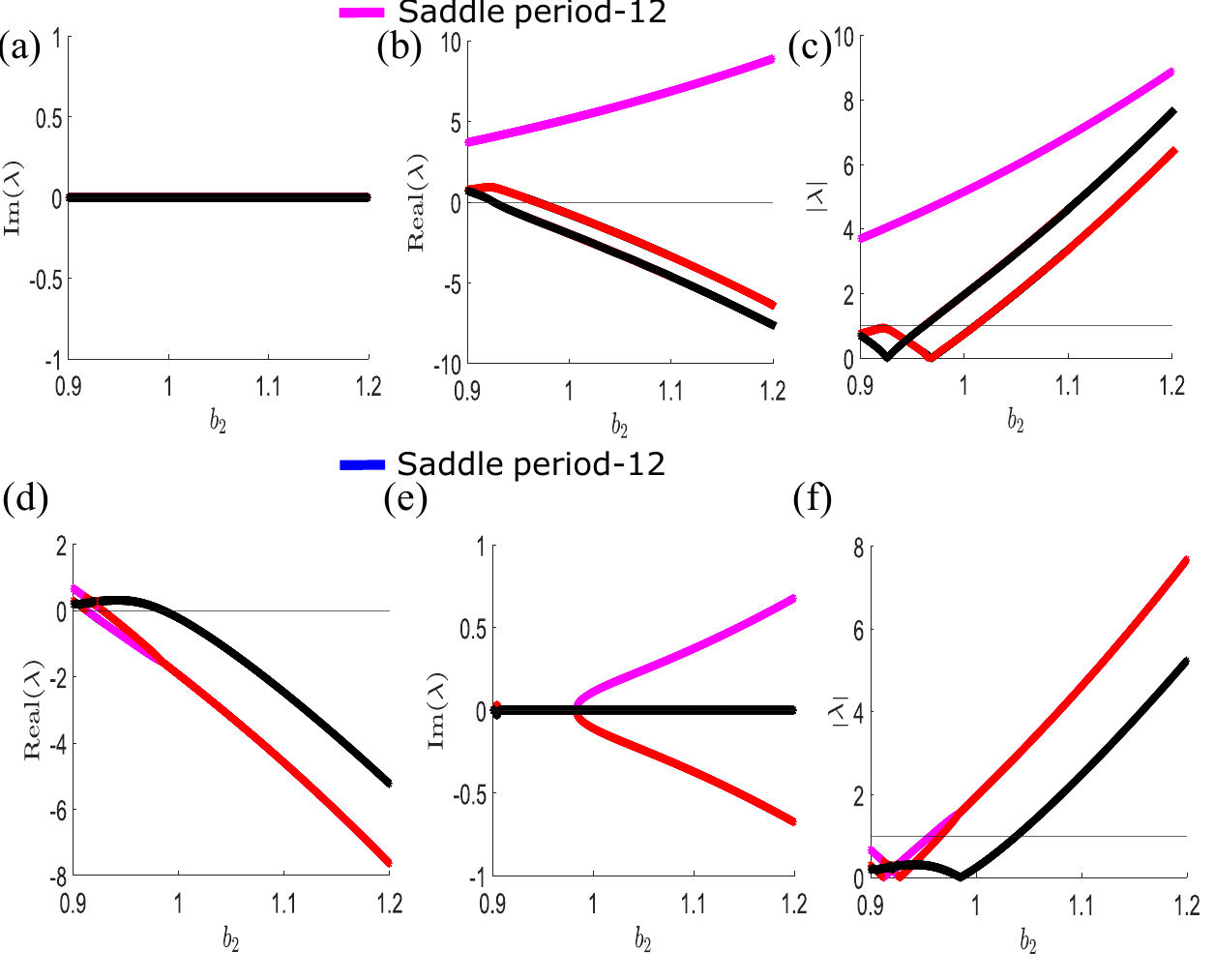}
\end{center}
\caption{One-parameter variation of the eigenvalues of the unstable period-12 orbit with respect to parameter $b_2$. In (a)-(c), the real, imaginary, and absolute part of the three eigenvalues of unstable period-12 orbit (marked in magenta) with $b_2$ is shown with different colours. In (d)-(f), the real, imaginary, and absolute part of the three eigenvalues of unstable period-12 orbit (marked in blue) with $b_2$ is shown. Observe that the nature of both the coexisting unstable period-12 orbits are different.}
\label{fig:Eigen_Saddlep12}
\end{figure*}
The variation of eigenvalues of two different saddle period-12 orbits are shown in Fig. \ref{fig:Eigen_Saddlep12}. For the saddle periodic orbit denoted in Magenta in Fig. \ref{fig:Eigen_Saddlep12}(a), all the eigenvalues are real with the imaginary part zero. As parameter $b_2$ increases, two real eigenvalues crosses unity continuously. For $b_2 > 1.1$, we observe three eigenvalues greater than unity in absolute value rendering the unstable invariant set to be a repelling period-12 orbit. 

For the saddle period-12 orbit denoted in blue in Fig. \ref{fig:STM_OneParam_Saddles} (a), the variation of eigenvalues with parameter $b_2$ are shown in Fig. \ref{fig:Eigen_Saddlep12} (d)- (f). As parameter $b_2$ increases, the eigenvalues continuously crosses unity. For $b_2 > 1.1$, we observe two eigenvalues are complex conjugate with magnitude greater than one and the third eigenvalue is real with modulus greater than one implying the unstable invariant set to be a repelling focus. 
\begin{figure*}[!htbp]
\hspace{2cm}\begin{center}
\includegraphics[width=0.9\textwidth]{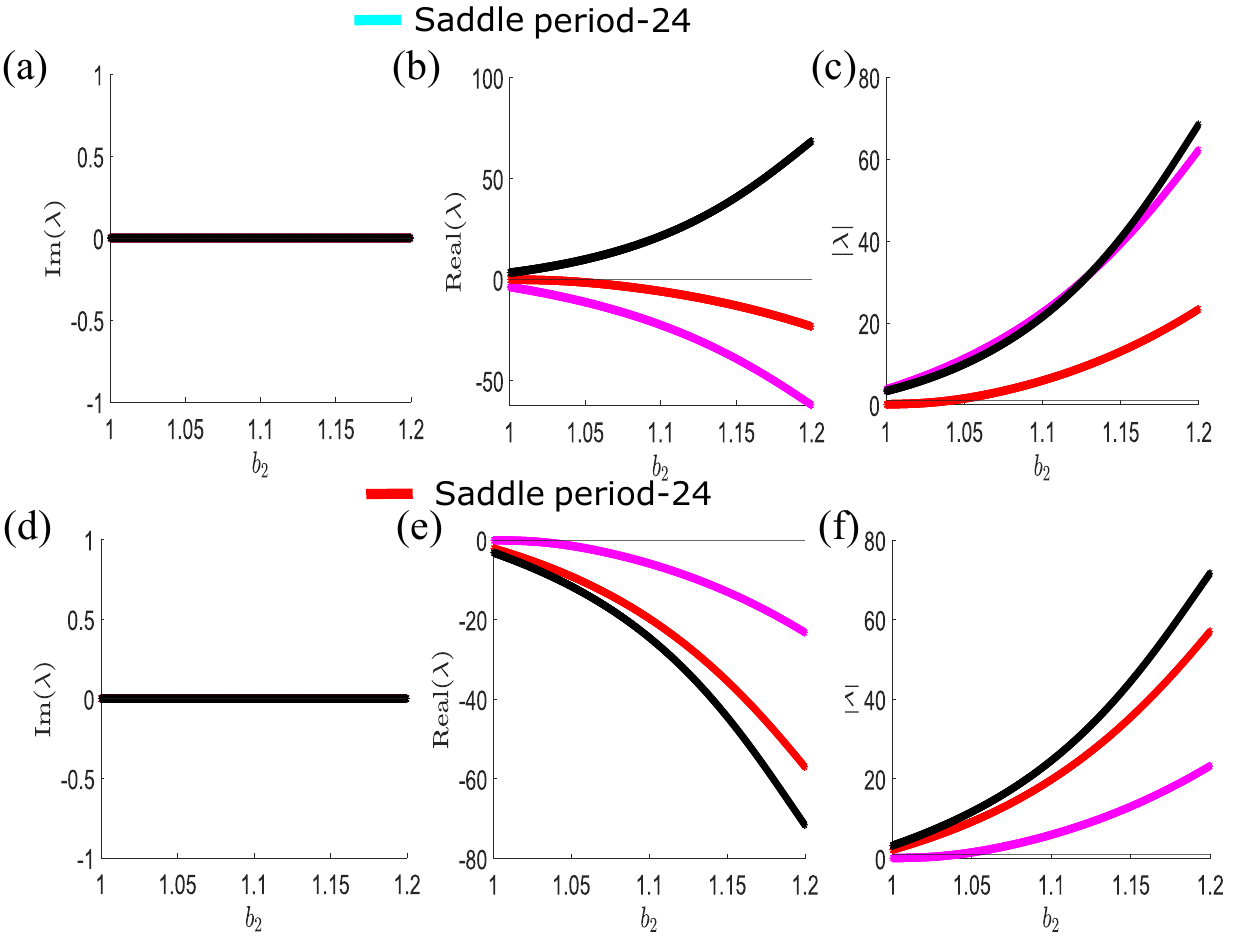}

\end{center}
\caption{One-parameter variation of the eigenvalues of the unstable period-24 orbit with respect to parameter $b_2$. In (a)-(c), the real, imaginary, and absolute part of the three eigenvalues of unstable period-24 orbit (marked in cyan) with $b_2$ is shown with different colours. In (d)-(f), the real, imaginary, and absolute part of the three eigenvalues of unstable period-24 orbit (marked in red) with $b_2$ is shown. }
\label{fig:Eigen_Saddlep24}
\end{figure*}
In Fig. \ref{fig:Eigen_Saddlep24}, variation of eigenvalues with $b_2$ are shown for different saddle period-24 periodic orbits (marked in cyan and red in Fig. \ref{fig:STM_OneParam_Saddles} (a)). The eigenvalues for both the unstable invariant sets are greater than unity and are all real implying both of them to be repelling saddle period-24 orbit. 

Observe that the transition from two positive Lyapunov exponents to three postive Lyapunov exponents takes pace when apart from the two eigenvalues with modulus greater than one, the third eigenvalue crosses unity in modulus. The coexistence of unstable periodic orbits with different number of unstable directions leads to a kind of nonhyperbolicity known in the literature as Unstable Dimensional Variability (UDV) \cite{UDV} and can be a possible mechanism for the continuous transition of Lyapunov exponents through zero.

\begin{figure*}[!htbp]
\hspace{-2cm}\begin{center}
\includegraphics[width=0.9\textwidth]{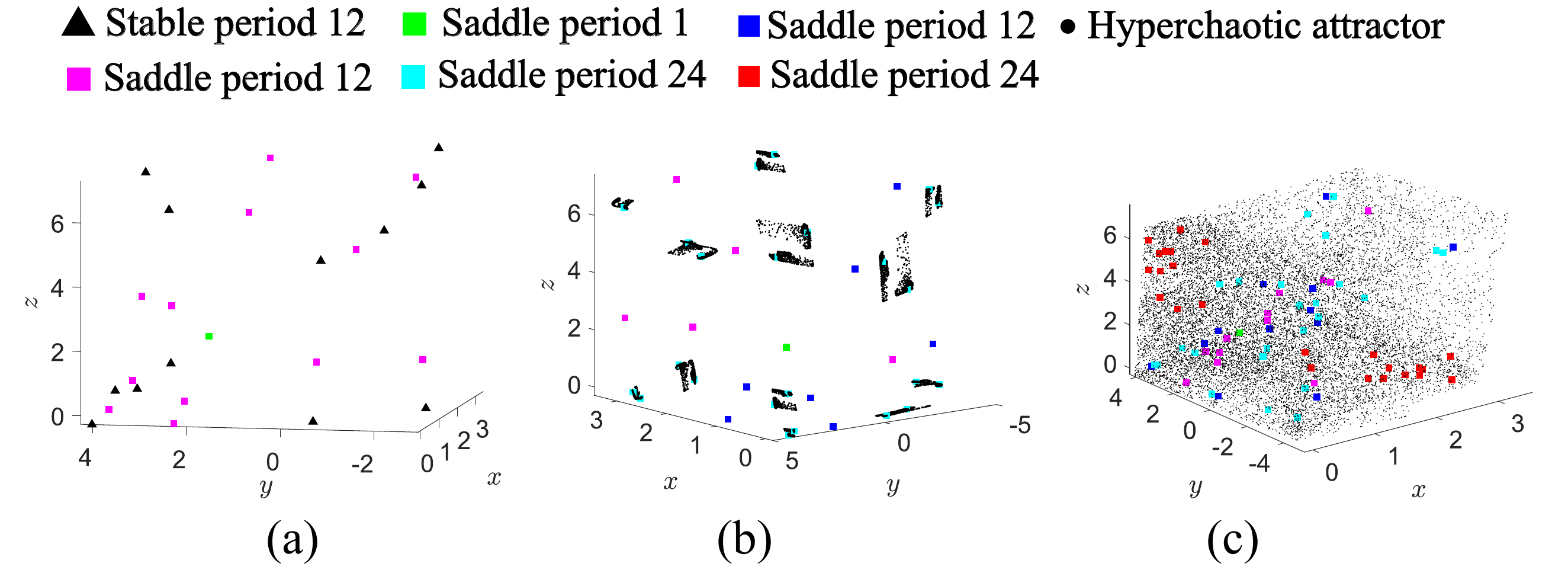}
\end{center}
\caption{Phase portraits of the unstable invariant sets and coexisting attractors. In (a), for $b_2 = 0.95$, a stable period-12 orbit (marked by black triangles) coexists with unstable period-12 orbit (marked by squares in magenta) and unstable fixed point (marked by square in green). In (b), for $b_2 = 0.99$, hyperchaotic attractor with 
two positive Lyapunov exponents are shown along with unstable fixed point (in green), unstable period-12 orbit (in magenta and blue), and unstable period-24 orbit (in cyan). Observe that the unstable period-24 orbit is absorbed in the hyperchaotic attractor. In (c), for $b_2 = 1.2$, hyperchaotic attractor with three positive Lyapunov exponents 
is shown which is in shape of a cube along with coexisting unstable fixed point, period-12 orbit, period-24 orbit, all of them are absorbed in the hyperchaotic attractor.}
\label{fig:Hyperchaotic_saddle}
\end{figure*}

After obtaining information about the eigenvalues of the unstable invariant sets 
coexisting in the regime of hyperchaos, we illustrate via phase portraits the absorption of different repelling orbits in the hyperchaotic attractor, see Fig. \ref{fig:Hyperchaotic_saddle}. In Fig. \ref{fig:Hyperchaotic_saddle}, at $b_2  = 0.95$, we show the coexistence of stable period-12 orbit (in black triangles), repelling-focus period-12 orbit, and repelling-focus saddle fixed point (in green squares). In Fig. \ref{fig:Hyperchaotic_saddle} (b) at $b_2 = 0.99$ we show sheet like hyperchaotic attractor with two positive Lyapunov exponents, and the coexistence of repelling focus period-12 orbit (in magneta squares), absorbed repelling period-24 orbit (in cyan), and repelling focus period-12 orbit in blue squares. The sheet like hyperchaotic attractor in black suggests that the expansion is occurring in two-dimensions. In Fig. \ref{fig:Hyperchaotic_saddle} (c), for $b_2 = 1.2$ shows a cube shaped hyperchaotic attractor with three positive Lyapunov exponents illustrating the expansion occuring in three dimensions, and absorbed repelling focus period-12 orbit (in magenta and blue), repelling period-24 orbit (in red and cyan), repelling focus fixed point (in green).

The next section deals with a route from doubling bifurcation of quasiperiodic closed invariant curves to hyperchaotic regime. 
\section{From ergodic doubling bifurcation to hyperchaos}
\label{sec:DoublingHyperchaos} 

\begin{figure}[!htbp]
\hspace{-2cm}\begin{center}
\includegraphics[width=0.7\textwidth]{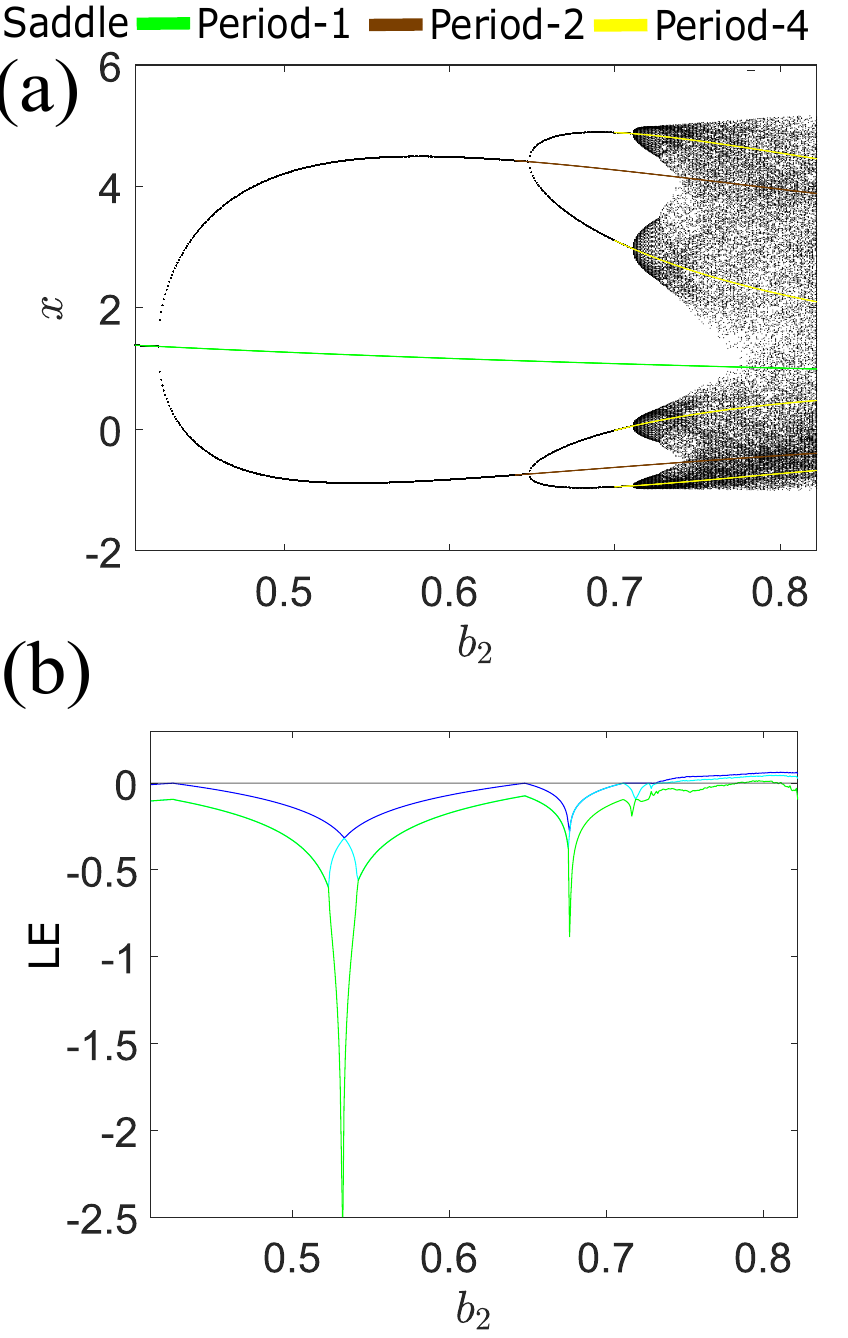}

\end{center}
\caption{In (a), One-parameter bifurcation of $x$ vs $b_2$ showing ergodic torus doubling bifurcation. Different unstable periodic orbits are continued as shown in different colours in the legend. In (b), variation of three Lyapunov exponents are shown.}
\label{fig:OneParamSaddles_Doubling}
\end{figure}

Fig. \ref{fig:OneParamSaddles_Doubling}(a) shows a one-parameter bifurcation diagram of $x$ vs $b_{2}$. A stable fixed point undergoes two subsequent period-doubling bifurcations, leading to the formation of 
stable period-four orbit. The stable period-four orbit then undergoes a supercritical Neimark-Sacker bifurcation leading to the formation of four disjoint cyclic quasiperiodic closed invariant curve. With increase in parameter $b_2$, a doubling 
bifurcation of quasiperiodic closed invariant curve takes place, a detailed discussion is made in \cite{Muni23b}. A length-doubled quasiperiodic closed invariant curve is born. As parameter $b_{2}$ is increased further, the closed invariant curve loses smoothness and transforms into a hyperchaotic attractor (see Fig \ref{fig:PhaseSpaceDoubling}(b)). For a small interval of $b_{2}$, there are three positive Lyapunov exponents. 

\begin{figure}[!htbp]
\hspace{-2cm}\begin{center}
\includegraphics[width=0.9\textwidth]{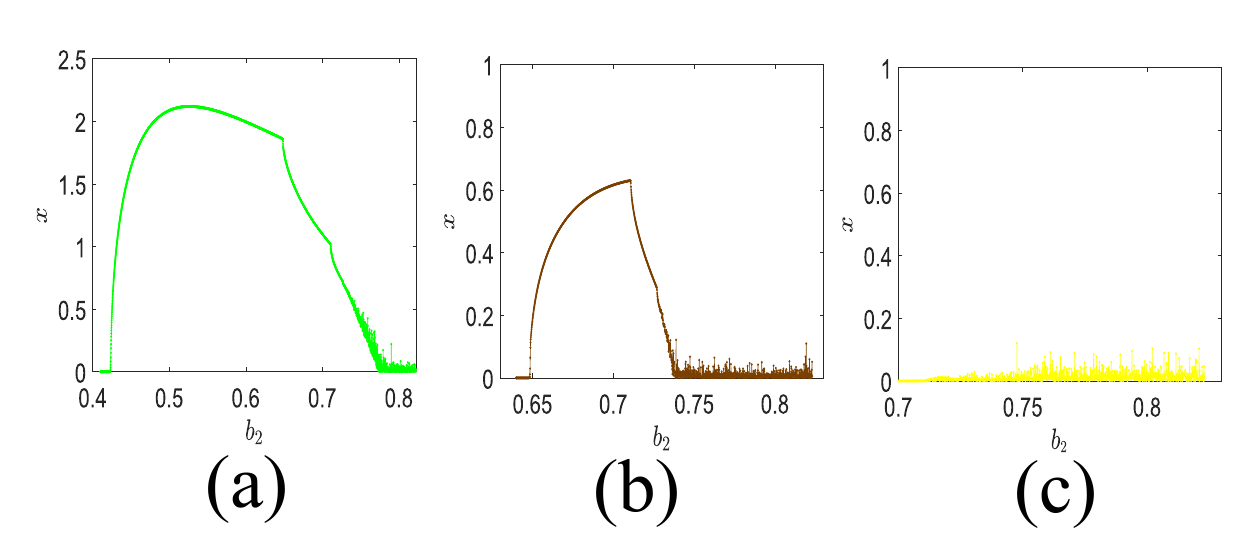}

\end{center}
\caption{Variation of the distance between the saddle fixed points, periodic points and the coexisting attractors. In (a), distance between the unstable fixed point and coexisitng attractors are shown in green. In (b), distance between unstable period-2 orbits (in brown ) and coexisting attractors. In (c), distance between unstable period-4 orbits and the coexisting attractors are shown in yellow. Parameter values at which the distance approaches zero suggest that the unstable invariant set is absorbed by the coexisting attractor at that parameter value.}
\label{fig:DistanceCollageDoubling}
\end{figure}

\begin{figure}[!htbp]
\hspace{-2cm}\begin{center}
\includegraphics[width=0.7\textwidth]{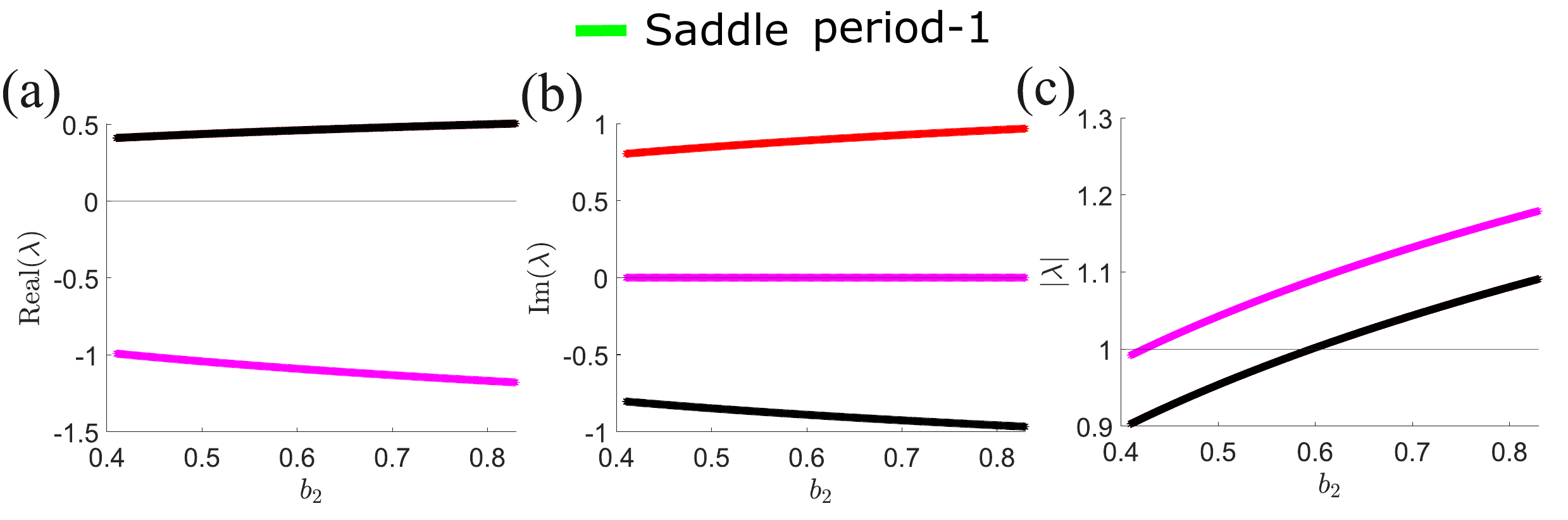}

\end{center}
\caption{Variation of the eigenvalues of the saddle fixed point with 
parameter $b_2$. In (a), the real part of the eigenvalue with respect to parameter is considered. In (b), the imaginary part 
of the eigenvalue with respect to parameter is shown. In (c), 
the magnitude of the eigenvalue with respect to parameter is 
considered. 
 }
\label{fig:EigenP1Doubling}
\end{figure}

\begin{figure}[!htbp]
\hspace{-2cm}\begin{center}
\includegraphics[width=0.6\textwidth]{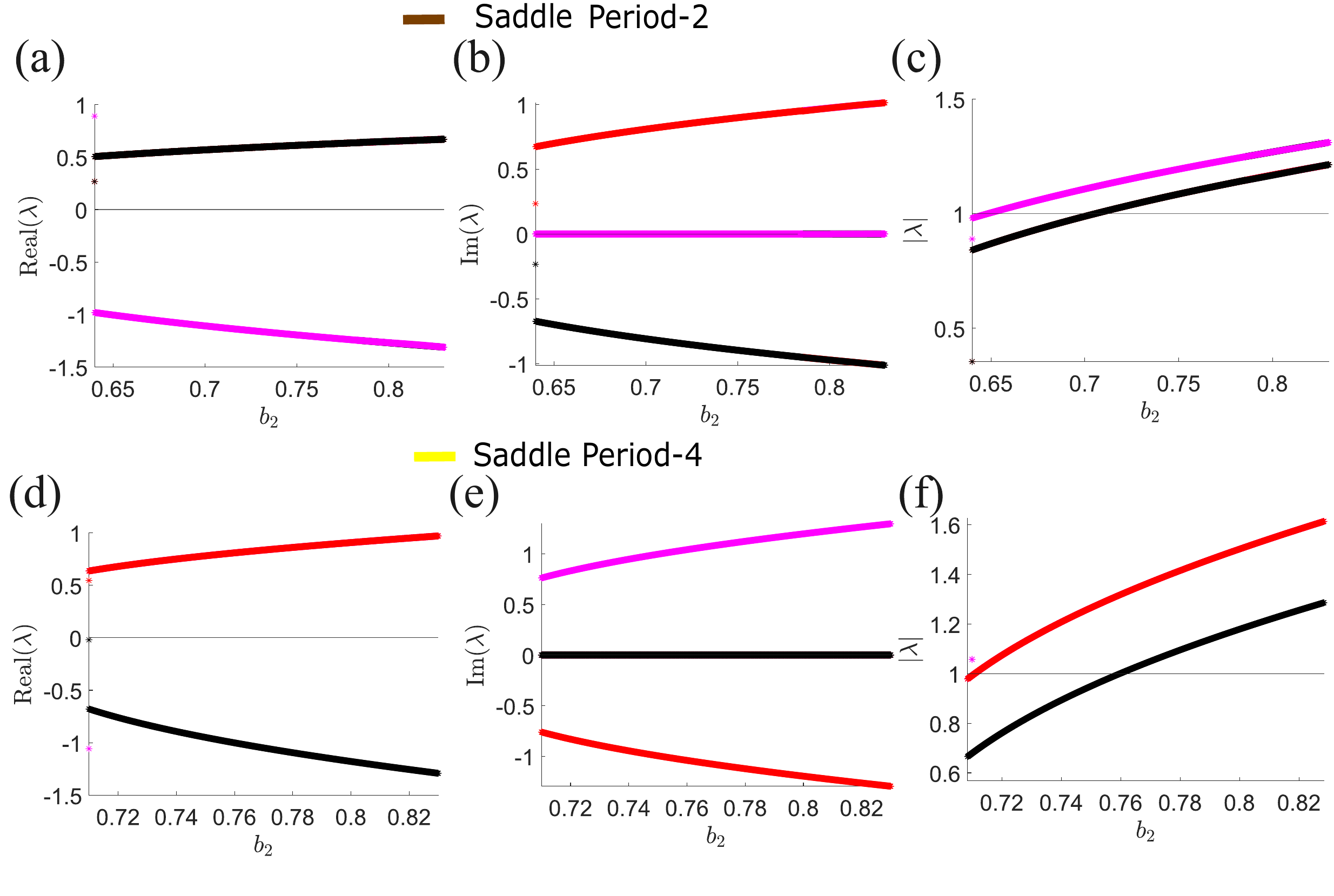}
\end{center}
\caption{ In (a)-(c), the real part, imaginary part, and magnitude of the eigenvalues of saddle period-two orbit with respect to parameter $b_2$ is considered. In (d)-(f), the real part, imaginary part, and magnitude of the eigenvalues of saddle period-four orbit with respect to parameter $b_2$ is considered. Observe that the trend and type of unstable orbits are similar, see Fig. \ref{fig:EigenP1Doubling}. 
}
\label{fig:EigenP2P4Doubling}
\end{figure}

\begin{figure}[!htbp]
\hspace{-2cm}\begin{center}
\includegraphics[width=0.7\textwidth]{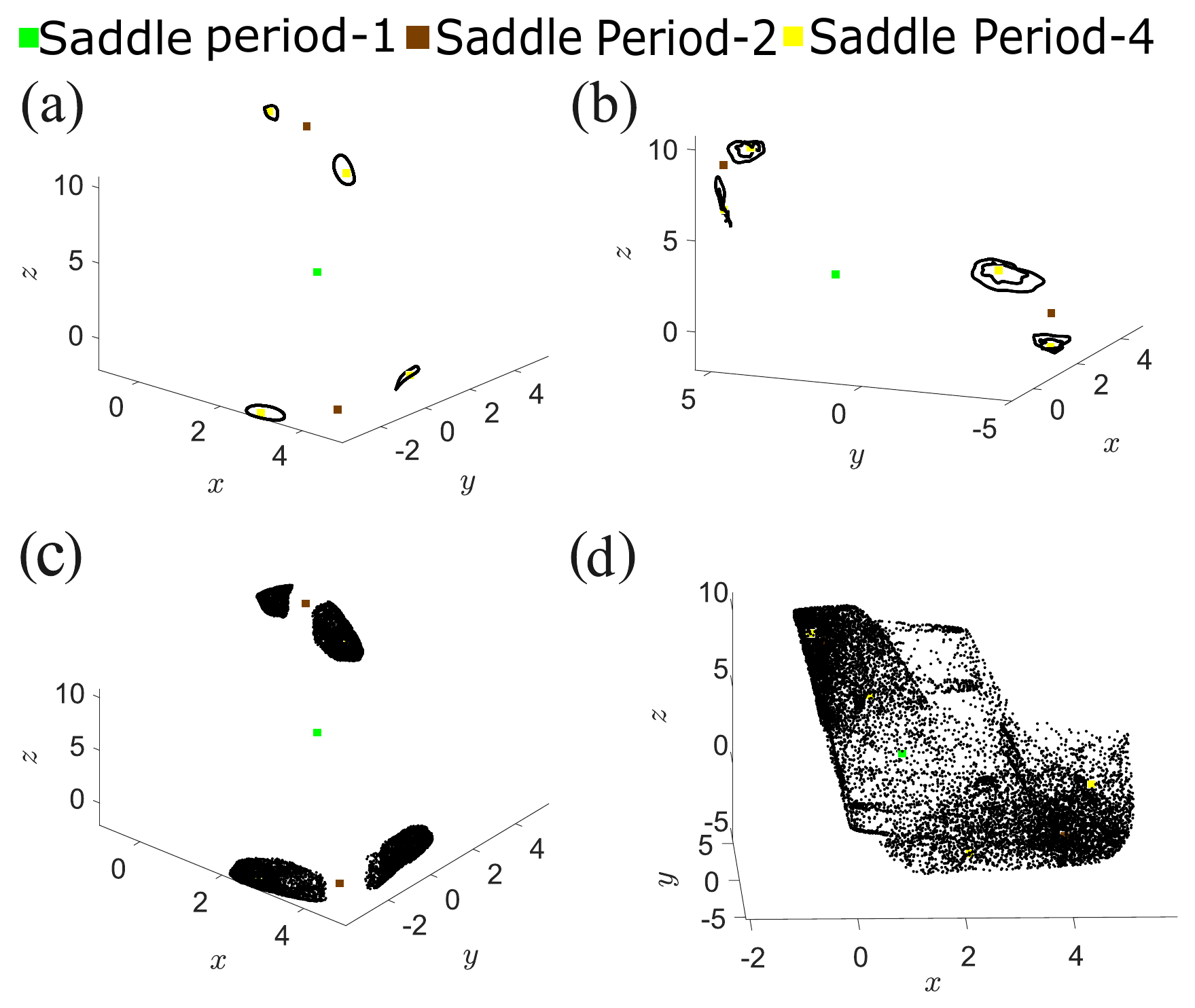}
\end{center}
\caption{Phase portraits with coexisting saddle periodic orbits. In (a), for $b_2 = 0.72$, coexistence of four cyclic closed invariant curve along with saddle period-two orbit (marked in brown) and saddle fixed point (marked in green). In (b), for $b_2 = 0.73$, the doubled closed invariant curve is shown. In (c), for $b_2 = 0.74$, hyperchaotic attractor with two positive Lyapunov exponents is shown along with unstable periodic orbits. In (d), for $b_2 = 0.82$, a hyperchaotic attractor is shown after an attractor merging crisis, absorbing all unstable periodic points present in the system.}
\label{fig:PhaseSpaceDoubling}
\end{figure}

We continuate the periodic orbits as shown in Fig. \ref{fig:OneParamSaddles_Doubling}(a) till the regime of hyperchaos in the system. The saddle fixed point is shown in green. The period-two orbit is marked in magenta, and the period-four saddle orbit is marked in yellow. We compute the distance between the saddle periodic orbits and the attractors which can give an idea about the absorbing domain in the hyperchaotic attractor. The distance computed are shown in Fig. 8. In Fig. \ref{fig:DistanceCollageDoubling}(a), distance between the saddle period-one orbit and the attractors are shown, see the for $b_{2}<0.8$, the distance is non-zero, showing a separation between the attractors. For nearby values of $b_{2} \approx 0.8$, the distance approaches zero, showing that the fixed point is absorbed inside the hyperchaotic attractor. In Fig. \ref{fig:EigenP1Doubling} (a) - (c), the real part, imaginary part, and magnitude of the eigenvalues are shown respectively. We note that the three eigenvalues are greater than one in abolute value (one real  and a pair of complex conjugate eigenvalues) in the regime of hyperchaos. We can also observe a similar behavior for the period-two (see Fig. \ref{fig:EigenP2P4Doubling} (a) - (c)) and period-four orbits (see Fig. \ref{fig:EigenP2P4Doubling}), which are of repelling nature with one real and a pair of complex conjugate eigenvalues. 

We show some selected phase portraits of the formation of hyperchaotic attractors along with the repelling invariant sets in Fig. \ref{fig:PhaseSpaceDoubling}. In Fig.\ref{fig:PhaseSpaceDoubling}(a), a disjoint four cyclic quasiperiodic closed invariant curve is observed. In Fig. \ref{fig:PhaseSpaceDoubling} (b), a length doubled disjoint quasiperiodic closed curve. \textcolor{black}{In (c), a four piece cyclic hyperchaotic attractor with two positive Lyapunov exponents $(-0.03214,0.00304,0.02264)$ is shown. In (d), formation of a hyperchaotic attractor with two positive Lyapunov exponents $(-0.04338,0.03898,0.05939)$ after an attractor merging crisis between the four cyclic hyperchaotic attractor takes place.} 

The next section illustrates the presence of a weak flow like hyperchaotic attractor in a narrow region of parameter space. 
\section{Weak hyperchaos: flow like hyperchaotic attractor}
\label{sec:weak_hyperchaos}
The next section deals with a weak hyperchaotic attractor in which a Lyapunov exponent is positive, second Lyapunov exponent is positive but close to zero, and the third Lyapunov exponent is negative.

\begin{figure}[!htbp]
\hspace{-2cm}\begin{center}
\includegraphics[width=0.6\textwidth]{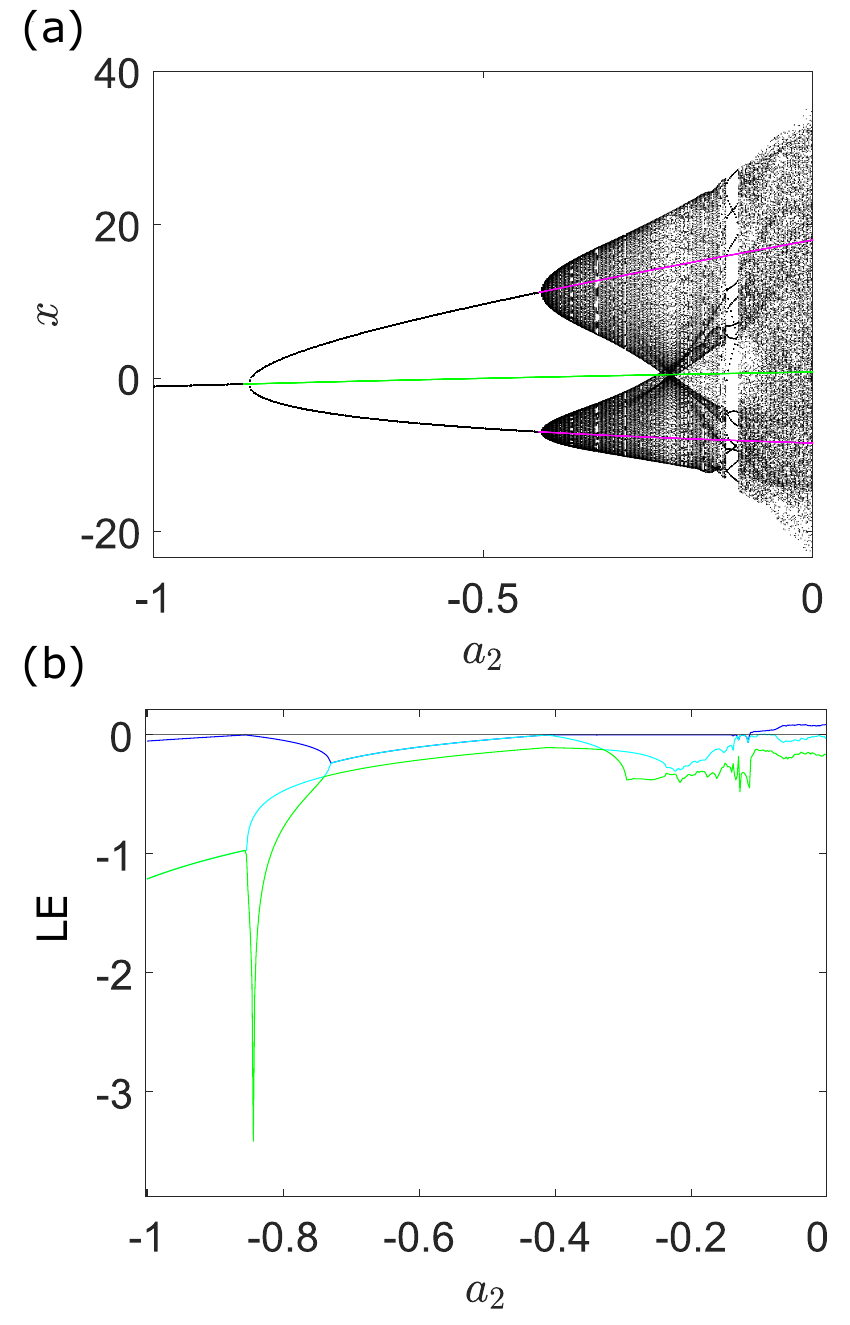}
\end{center}
\caption{In (a), a one-parameter bifurcation diagram with respect to parameter $a_2$ is shown along with three Lyapunov exponents 
marked in different colour in (b). The parameters are $a_1 = -0.856273, a_3 = 0.12, b_1 = 4, b_2=0.3, c=2.15$.}
\label{fig:a2_OneParam}
\end{figure}

Fig. \ref{fig:a2_OneParam} (a) shows a one-parameter bifurcation diagram of $x$ vs $a_{2}$. At $a_{2} = -1$, there exists a stable fixed point of the system. As $a_{2}$ increases, it undergoes a period-doubling bifurcation. With further increase in $a_{2}$, they undergo a supercritical Neimark-Sacker bifurcation leading to the formation of two cyclic disjoint quasiperiodic invariant curves. 
After an attractor merging crisis, the two quasiperiodic 
closed invariant curve merge and transform to a larger quasiperiodic closed invariant curve. With increase in $a_{2}$, 
a reverse saddle-node bifurcation on an invariant circle takes place, leading to the formation of a mode-locked period-17 orbit. This periodic orbit undergoes a boundary crisis which leads to the formation of weak hyperchaotic attractor (with near zero positive Lyapunov exponent).  In Fig. \ref{fig:a2_OneParam}, we also continuate a saddle fixed point and a stable period-two point shown in green and magenta respectively.

\begin{figure}[!htbp]
\hspace{-2cm}\begin{center}
\includegraphics[width=0.7\textwidth]{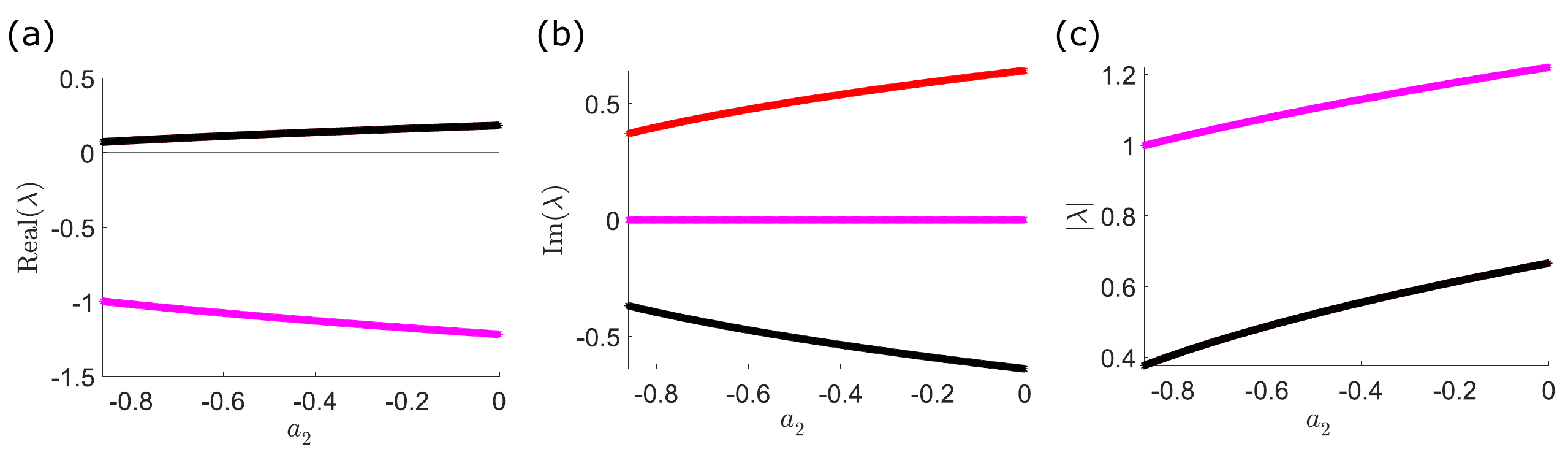}
\end{center}
\caption{Variation of eigenvalues of unstable fixed point with parameter $a_2$. In (a), real part of eigenvalue is shown. In (b), imaginary part of eigenvalue of unstable fixed point is shown. In (c), modulus of the eigenvalue is shown with variation of parameter $a_2$. The parameters are $a_1 = -0.856273, a_3 = 0.12, b_1 = 4, b_2=0.3, c=2.15$.}
\label{fig:a2_Saddles}
\end{figure}

In Fig. \ref{fig:a2_Saddles}, the real parts, imaginary part, and absolute value of the eigenvalue of the saddle fixed point are shown. At the hyperchaotic regime, we have one real eigenvalue with modulus greater than one and a pair of complex conjugate eigenvalues with modulus greater than one. A repelling focus fixed point is absorbed inside the weak flow like hyperchaotic attractor. A similar nature of eigenvalues are observed for other periodic points.

\begin{figure}[!htbp]
\hspace{-2cm}\begin{center}
\includegraphics[width=0.7\textwidth]{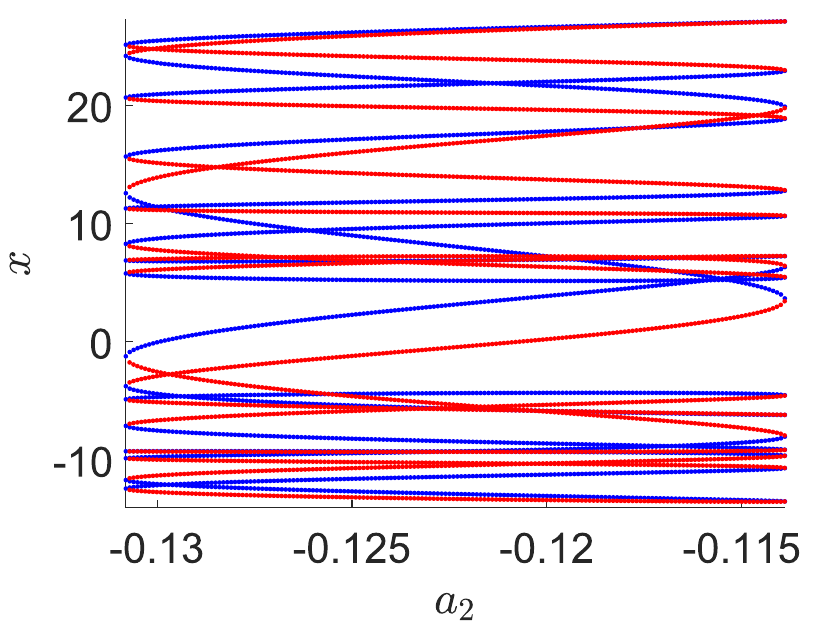}
\end{center}
\caption{A one-parameter continuation of stable and saddle period-17 orbit to show the existence of a mode-locked period-$17$ orbit. The curves in blue denote the stable period-17 orbit and in red denotes the saddle 
period-17 orbit. Observe the occurrence of saddle-node bifurcation. The parameters are $a_1 = -0.856273, a_3 = 0.12, b_1 = 4, b_2=0.3, c=2.15$.}
\label{fig:a2_SaddleNodeP17}
\end{figure}

For better illustration, we have shown separately in Fig. \ref{fig:a2_SaddleNodeP17} the region of mode-locked periodic orbits. The stable and saddle periodic orbits are shown in blue and red respectively. All seventeen points both saddle and stable orbits are shown. 

\begin{figure}[!htbp]
\hspace{-2cm}\begin{center}
\includegraphics[width=0.7\textwidth]{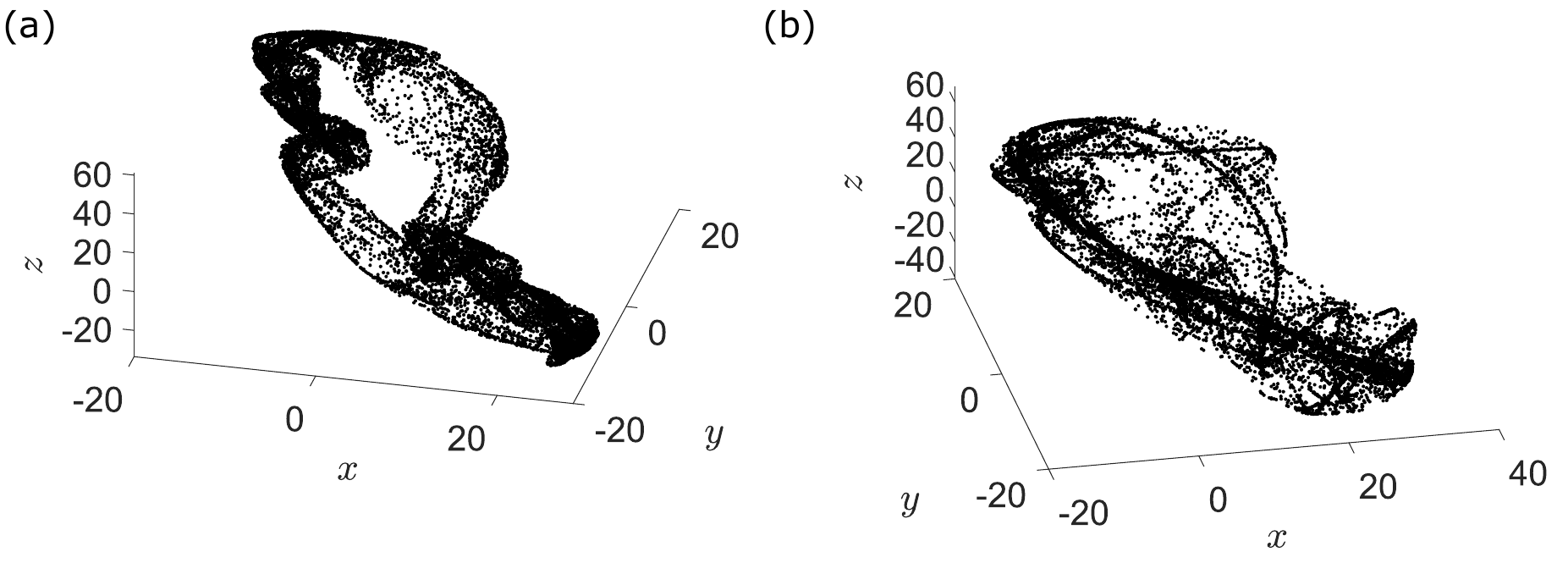}
\end{center}
\caption{In (a) for $a_2 = -0.108$, weak flow like hyperchaotic attractor is shown. In (b) $a_2 = -0.048$, a chaotic attractor is shown. The parameters are $a_1 = -0.856273, a_3 = 0.12, b_1 = 4, b_2=0.3, c=2.15$.}
\label{fig:a2_Phaseportrait}
\end{figure}

In Fig. \ref{fig:a2_Phaseportrait} (a), for $a_{2} = -0.108$, a weak hyperchaotic attractor with two positive Lyapunov exponents are shown. As $a_{2}$ increases further, in Fig. \ref{fig:a2_Phaseportrait}(b), for $a_{2} = -0.048$, a chaotic attractor is shown.  

\section*{Conclusions}
Previous studies were made for hyperchaotic attractors in three-dimensional maps with two positive Lyapunov exponents. In this paper, a three-dimensional map under consideration displays strong hyperchaos in the sense that there exists a wide parameter regime in which all three Lyapunov exponents are positive. We investigate the presence of saddle periodic points absorbed in the hyperchaotic attractors. This is carried out by computing the Euclidean distance of the saddle periodic points and the hyperchaotic attractors. It has been discussed in \cite{Muni23b} that the map displays both ergodic and resonant torus doublings. As discussed in \cite{Shykhmamedov_2023}, the doublings of quasiperiodic closed invariant curves is one of the routes to the formation of Shilnikov attractors. In this paper, after subsequent doublings of quasiperiodic closed invariant curve, we also illustrate the existence of a hyperchaotic attractor with two positive Lyapunov exponents and the unstable periodic points absorbing them have three eigenvalues with absolute values greater than unity. Two-dimensional Lyapunov charts are made to discuss and identify the saddle periodic points under two-parameter continuation.  It would be interesting to verify whether the points in the cube hyperchaotic attractors fill it uniformly. Moreover understanding on the mechanism of the transition from two positive Lyapunov exponents to three positive Lyapunov exponents would be interesting. \textcolor{black}{In this article we have shown the role played by
the saddle periodic orbits in their absorption by hyperchaotic attractors. The computation of saddle and one-dimensional manifolds are performed by numerical methods and are analytically intractable. We have seen that the dimensionality of the unstable manifold of the saddle periodic orbit matches with the number of positive Lyapunov exponents of the hyperchaotic attractor. This numerical observation needs to be further strengthened using theoretical arguments and a generalised theorem can be established in future works.}

Moreover, we have shown that repelling periodic points are absorbed in the hyperchaotic attractors with all three positive Lyapunov exponents. It remains to check whether the repelling periodic points are snap-back repellers or not? It would be interesting to develop a control strategy to control the extreme hyperchaotic behaviors which can be useful in some real experiments. \textcolor{black}{It would be interesting to consider the electronic realization of the three dimensional discrete quadratic map and possible application of secure communication. The 3D quadratic map considered in this study can be implemented experimentally in electronic circuit due to the presence of simple quadratic terms in the map along similar lines of \cite{Fu2023}. The hyperchaotic behavior of the output signal could be validated by computing Lyapunov exponents from the data.
}
\section*{Acknowledgements}
	S.S.M acknowledges insightful discussions with Alexey Kazakov on two-parameter continuation techniques.   
\section*{Conflict of interest}
 The authors declare that they have no conflict of interest.
\section*{Data Availability Statement}
 The data that support the findings of this study are available within the article.


\bibliographystyle{unsrt}
\bibliography{ReferHyperchaos}

\begin{thebibliography}{10}

\bibitem{Rossler79}
O.E. Rossler.
\newblock An equation for hyperchaos.
\newblock {\em Physics Letters A}, 71(2):155--157, 1979.

\bibitem{UDV}
R.L. Viana and C.~Grebogi.
\newblock Riddled bains and unstable dimension variability in chaotic systems with and without symmetry.
\newblock {\em International Journal of Bifurcation and Chaos}, 11(10):2689--2698, 2001.

\bibitem{Muni23a}
S.S. Muni.
\newblock Mode-locked orbits, doubling of invariant curves in discrete {H}indmarsh-{R}ose neuron model.
\newblock {\em Physica Scripta}, 2023.

\bibitem{ModeLocked23}
S.~S. Muni and S.~Banerjee.
\newblock Bifurcations of mode-locked periodic orbits in three-dimensional maps.
\newblock {\em International Journal of Bifurcation and Chaos}, 33(10):2330025, 2023.

\bibitem{Muni23b}
S.S. Muni.
\newblock Ergodic and resonant torus doubling bifurcation in a three-dimensional discrete quadratic map.
\newblock {\em arXiv}, 2023.

\bibitem{Shykhmamedov_2023}
A.~Shykhmamedov, E.~Karatetskaia, A.~Kazakov, and N.~Stankevich.
\newblock Scenarios for the creation of hyperchaotic attractors in 3d maps.
\newblock {\em Nonlinearity}, 36(7):3501, may 2023.

\bibitem{Ren_2017}
C.~Ren, J.~Zhou, and C.~Liu.
\newblock Chaos control of a multi-dimensional chaotic mapping system by modified stability transformation method.
\newblock {\em Journal of Vibroengineering}, 19(2):1103--1115, mar 2017.

\bibitem{KapitiniakRiddling}
T.~Kapitaniak.
\newblock Chaos synchronization and hyperchaos.
\newblock {\em Journal of Physics: Conference Series}, 23(1):317, jan 2005.

\bibitem{Kapit95}
I.~Cohent T.~Kapitaniak, K.E.~Thylew and J.~Wojewoda.
\newblock Chaos-hyperchaos transition.
\newblock {\em Chaos, Solitons \& Fractals}, 5(10):2003--2011, 1995.

\bibitem{SquidHyper21}
J.~Shena, N.~Lazarides, and J.~Hizanidis.
\newblock {Synchronization transitions in a hyperchaotic SQUID trimer}.
\newblock {\em Chaos: An Interdisciplinary Journal of Nonlinear Science}, 31(9):093102, 09 2021.

\bibitem{HyperApp1}
Song Z.
\newblock Synchronization analysis of complex-variable chaotic systems with discontinuous unidirectional coupling.
\newblock {\em Complexity}, 21(6):343--355, 2016.

\bibitem{HyperApp2}
L.~Munteanu, C.~Brişan, and V.~Chiroiu.
\newblock Chaos–hyperchaos transition in a class of models governed by sommerfeld effect.
\newblock {\em Nonlinear Dyn}, 78:1877–1889, 2014.

\bibitem{ShilnikovFocus}
I.~M. Ovsyannikov and L.~P. Shil'nikov.
\newblock On systems with a saddle-focus homoclinic curve.
\newblock {\em Mathematics of the USSR-Sbornik}, 58(2):557, feb 1987.

\bibitem{Kara21}
E.~Karatetskaia, A.~Shykhmamedov, and A.~Kazakov.
\newblock {S}hilnikov attractors in three-dimensional orientation-reversing maps.
\newblock {\em Chaos: An Interdisciplinary Journal of Nonlinear Science}, 31(1):011102, 2021.

\bibitem{KapitiniakChua}
T.~Kapitaniak, L.~Chua, and G.~Q. Zhong.
\newblock Experimental hyperchaos in coupled chua's circuits.
\newblock {\em IEEE Transactions on Circuits and Systems I-regular Papers}, 41:499--503, 1994.

\bibitem{FluidHyperchaos}
G.~C. Layek and N.~C. Pati.
\newblock Bifurcations and hyperchaos in magnetoconvection of non-newtonian fluids.
\newblock {\em International Journal of Bifurcation and Chaos}, 28(10):1830034, 2018.

\bibitem{HyperWater}
Frederick~D. Tappert, Gustavo~J. Goni, and Michael~J. Brown.
\newblock {Chaos and hyperchaos in shallow water acoustics}.
\newblock {\em The Journal of the Acoustical Society of America}, 84(S1):S152--S152, 08 2005.

\bibitem{Elw99}
A.S. Elwakil and M.P. Kennedy.
\newblock Inductorless hyperchaos generator.
\newblock {\em Microelectronics Journal}, 30(8):739--743, 1999.

\bibitem{Koc92}
Lj. Kocarev, K.~S. Halle, K.~Eckert, L.~O. Chua, and U.~Parlitz.
\newblock Experimental demonstration of secure communications via chaotic synchronization.
\newblock {\em International Journal of Bifurcation and Chaos}, 02(03):709--713, 1992.

\bibitem{Yassen2008}
M.T. Yassen.
\newblock Synchronization hyperchaos of hyperchaotic systems.
\newblock {\em Chaos, Solitons \& Fractals}, 37(2):465–475, July 2008.

\bibitem{Naderi2016}
B.~Naderi and H.~Kheiri.
\newblock Exponential synchronization of chaotic system and application in secure communication.
\newblock {\em Optik}, 127(5):2407–2412, March 2016.

\bibitem{10374589}
Kevin H.~M. Gularte, Felipe~O. Hara, José A.~R. Vargas, and Fábio~Oliveira Guimarães.
\newblock Hyperchaos-based secure communication using lyapunov theory.
\newblock In {\em 2023 15th IEEE International Conference on Industry Applications (INDUSCON)}, pages 747--751, 2023.

\bibitem{915393}
D.A. Miller and G.~Grassi.
\newblock Experimental realization of observer-based hyperchaos synchronization.
\newblock {\em IEEE Transactions on Circuits and Systems I: Fundamental Theory and Applications}, 48(3):366--374, 2001.

\bibitem{Gao08}
T.~Gao and Z.~Chen.
\newblock A new image encryption algorithm based on hyper-chaos.
\newblock {\em Physics Letters A}, 372(4):394--400, 2008.

\bibitem{Yang20}
F.~Yang, J.~Mou, J.~Liu, C.~Ma, and H.~Yan.
\newblock Characteristic analysis of the fractional-order hyperchaotic complex system and its image encryption application.
\newblock {\em Signal Processing}, 169:107373, 2020.

\bibitem{Natiq18}
H.~Natiq, Al-Saidi, Said N.M.G., and M.R.M.
\newblock A new hyperchaotic map and its application for image encryption.
\newblock {\em Eur. Phys. J. Plus}, 6, 2018.

\bibitem{Ozkaynak12}
F.~Özkaynak, A.~Bedri Özer, and S.~Yavuz.
\newblock Cryptanalysis of a novel image encryption scheme based on improved hyperchaotic sequences.
\newblock {\em Optics Communications}, 285(24):4946--4948, 2012.

\bibitem{Li19}
C.~Li, F.~Zhao, C.~Liu, L.~Lei, and J.~Zhang.
\newblock A hyperchaotic color image encryption algorithm and security analysis.
\newblock {\em Security and Communication Networks}, 2019.

\bibitem{Hu23}
X.~Hu, D.~Jiang, and M.~et~al. Ahmad.
\newblock Novel 3-d hyperchaotic map with hidden attractor and its application in meaningful image encryption.
\newblock {\em Nonlinear Dyn}, 2023.

\bibitem{Lai23a}
Qiang Lai, Liang Yang, and Guanrong Chen.
\newblock Design and performance analysis of discrete memristive hyperchaotic systems with stuffed cube attractors and ultraboosting behaviors.
\newblock {\em IEEE Transactions on Industrial Electronics}, pages 1--10, 2023.

\bibitem{Lai23b}
Q.~Lai, Y.~Liu, and L.~Yang.
\newblock Image encryption using memristive hyperchaos.
\newblock {\em Appl Intell}, 2023.

\bibitem{Wang23}
M.~Wang, M.~An, S.~He, X.~Zhang, H~Ho-Ching~Iu, and Z.~Li.
\newblock {Two-dimensional memristive hyperchaotic maps with different coupling frames and its hardware implementation}.
\newblock {\em Chaos: An Interdisciplinary Journal of Nonlinear Science}, 33(7):073129, 07 2023.

\bibitem{Zhang23}
Y~Zhang, Z~Hua, H~Bao, H~Huang, and Y~Zhou.
\newblock Generation of $n$ -dimensional hyperchaotic maps using gershgorin-type theorem and its application.
\newblock {\em IEEE Transactions on Systems, Man, and Cybernetics: Systems}, pages 1--14, 2023.

\bibitem{Liu23}
W~Liu, K~Sun, S~He, and H~Wang.
\newblock The parallel chaotification map and its application.
\newblock {\em IEEE Transactions on Circuits and Systems I: Regular Papers}, 70(9):3689--3698, 2023.

\bibitem{GarcaMartnez2013}
M.~García-Martínez, I.~Campos-Cantón, E.~Campos-Cantón, and S.~Čelikovský.
\newblock Difference map and its electronic circuit realization.
\newblock {\em Nonlinear Dynamics}, 74(3):819–830, August 2013.

\bibitem{Huang2005}
Yan Huang and Xiao-Song Yang.
\newblock Chaoticity of some chemical attractors: a computer assisted proof.
\newblock {\em Journal of Mathematical Chemistry}, 38(1):107–117, July 2005.

\bibitem{QRmethod}
H.~F. {von Bremen}, F.~E. Udwadia, and W.~Proskurowski.
\newblock An efficient qr based method for the computation of lyapunov exponents.
\newblock {\em Physica D: Nonlinear Phenomena}, 101(1):1--16, 1997.

\bibitem{KuMe19}
Y.A. Kuznetsov and H.G.E.
\newblock {\em Numerical Bifurcation Analysis of Maps: From Theory to Software}.
\newblock Cambridge Monographs on Applied and Computational Mathematics. Cambridge University press, 2019.

\bibitem{Fu2023}
S.~Fu, X.~Cheng, and J.~Liu.
\newblock Dynamics, circuit design, feedback control of a new hyperchaotic system and its application in audio encryption.
\newblock {\em Scientific Reports}, 13(1), November 2023.

\end{thebibliography}

\end{document}